\documentclass[traditabstract]{aa}
\usepackage[varg]{txfonts}
\usepackage{graphicx}
\usepackage{subfigure}
\usepackage{rotating}
\usepackage{xcolor}
\usepackage{svg}
\usepackage{isomath}
\usepackage{natbib}
\usepackage{ragged2e}
\usepackage{soul} % to bar text in the correction
\usepackage{hyperref}
\usepackage{float}
\usepackage{amsmath}
\usepackage{wrapfig,lipsum}
\usepackage[flushleft]{threeparttable} % to have tables with title and notes below
\usepackage{multirow}
\DeclareUnicodeCharacter{2212}{-}
\usepackage{tikz}
\usepackage{esvect}
\usepackage{siunitx}
\newcommand{\kms}{km\,s$^{-1}$}

\begin{document}

\title{Central kiloparsec region of Andromeda. I. Dynamical modeling}

 \author{ Lucie Cros
           \inst{1,2}
           \and
           Fran\c coise Combes\inst{2}\fnmsep\inst{3}
           \and
           Anne-Laure Melchior\inst{2}
           \and 
           Thomas Martin\inst{4,5},
           }
        \institute{Laboratoire de Physique de l’\'Ecole Normale Sup\'erieure, ENS, Universit\'e PSL, CNRS, Sorbonne Universit\'e, Universit\'e Paris Cit\'e, F-75005, Paris, France \\
              \email{lucie.cros@phys.ens.fr}
        \and
            LUX, Observatoire de Paris, Université PSL, Sorbonne Université, CNRS, 75005 Paris, France
        \and
             Coll\`{e}ge de France, 11 Place Marcelin Berthelot, 75005 Paris, France
        \and 
            D{\'e}partement de physique, de g{\'e}nie physique et d'optique, Universit{\'e} Laval, Qu{\'e}bec (QC), G1V 0A6, Canada
         \and 
            Centre de recherche en astrophysique du Qu{\'e}bec, Canada
 }

% \DeclareRobustCommand{\HII}{\textup{H\,\textsc{\lowercase{II}}}}
% \DeclareRobustCommand{\NII}{\textup{N\,\textsc{\lowercase{II}}}}
% \DeclareRobustCommand{\SII}{\textup{S\,\textsc{\lowercase{II}}}}
%         \DeclareRobustCommand{\OIII}{\textup{O\,\textsc{\lowercase{III}}}}
% \DeclareRobustCommand{\OII}{\textup{O\,\textsc{\lowercase{II}}}}
% \DeclareRobustCommand{\Halpha}{\textup{H$\alpha$}}
% \DeclareRobustCommand{\kms}{km\,s$^{-1}$}
% \DeclareRobustCommand{\cm1}{cm$^{-1}$}
% \DeclareRobustCommand{\degrees}{$^\circ$}
\newcommand{\Modif}[1]{\textcolor{purple}{#1}}
\date{Received 19 November 2024 / Accepted 6 February 2025}

% % \abstract{}{}{}{}{} 
% % 5 {} token are mandatory

\abstract{ The Andromeda galaxy (M31) is the nearest giant spiral galaxy to our own, which offers an opportunity to study dynamical phenomena occurring in nuclear disks and bulges at high resolution to explain star formation quenching and galaxy evolution through collisions and tides. Multi-wavelength data have revealed strong dynamical perturbations in the central kiloparsec (kpc) region of M31, with an off-centered tilted disk and ring, coinciding with a dearth of atomic and molecular gas. Our goal is to understand the origin of these perturbations and, thus, we propose a dynamical model that reproduces the global features of the observations. We report on the integral field spectroscopy of the ionized gas with H$\alpha$ and [NII] obtained with the Spectrom\`etre Imageur à Transform\'ee de Fourier pour l'\'Etude en Long et en Large de raies d'\' Emission (SITELLE), which is the optical imaging Fourier transform spectrometer (IFTS) at the Canada France Hawaii telescope (CFHT).  Using the fully sampled velocity field of ionized gas, together with the more patchy molecular gas velocity field previously obtained with the CO lines at IRAM-30m telescope and the dust photometry, we identified three dynamical components in the gas: the main disk, a tilted ring, and a nuclear warped disk. We computed a mass model for the central kpc, essentially from the stellar nuclear disk and  bulge, with minimal contributions from the main stellar and gaseous disk, along with a dark matter halo. The kinematics of the ionized and molecular gas was then computed in this potential, and the velocity field confronted qualitatively to observations. The best fit helped us determine the physical parameters of the three identified gas components: size, morphology, and geometrical orientation. These results are qualitatively compatible with a recent head-on collision with a M-32 like galaxy, as previously proposed. The kinematical observations correspond to a dynamical re-orientation of the perturbed nuclear disk, through a series of warps and tearing of the disk into the ring, following the collision.}
 \maketitle
\section{Introduction}
\label{sect:intro}
Andromeda is an SA(s)b galaxy characterized by an unusual morphology \citep{1964Sci...145..952A,1998A&A...338L..33H,2003ApJS..145..259H}. It exhibits weak spiral structures \citep[e.g.,][]{2006A&A...453..459N,2006AJ....131.2766R}, scarce star formation $SFR \simeq 0.4 M_\odot \,{yr}^{-1}$ concentrated at 10~kiloparsecs (kpc) in the main disk \citep[e.g.,][]{2006ApJ...650L..45B,Braun2009,Chemin2009,2010A&A...517A..77T,2011AJ....142..139A}, and two rings at 1~kpc and 10~kpc, as observed in the gas, dust, and star formation distributions \citep[e.g.,][]{2006ApJ...650L..45B,2006A&A...453..459N,2010A&A...517A..77T,2011AJ....142..139A}. It also features a central gas hole, long suspected to be of dynamical origin \citep{1992A&A...255..105J}. In addition, its non-symmetry observed at all scales both in the gas and stellar components \citep[e.g.,][]{1976A&A....50..421P,1992A&A...255..105J} is usually understood as being due to an intense collisional past, as supported by the presence of numerous relics observed as spectacular giant stellar loops and tidal streams in the outskirts \citep[e.g.,][]{2005ApJ...634..287I,2009Natur.461...66M,2018ApJ...868...55M}. 

Relying on the spectacular dust rings and spiral arms observed in the mid-infrared with the Infrared Array Camera (IRAC) on board the Spitzer Space Telescope \citep{2006ApJ...650L..45B}, \citet{Block2006} stressed the presence of an elongated off-center inner ring with projected diameters 1.5~kpc by 1~kpc. Given the fact that both rings at 1~kpc and 10~kpc are off-centered, \citet{Block2006} argued that the most likely scenario for the formation of these rings is a head-on collision of the Cartwheel type \citep{1993ApJ...411..108S,2001Ap&SS.276.1141H}. Unlike the Cartwheel, where the companion is about one-third of the mass of the target (major merger), in Andromeda, the collision can be called a minor merger, producing much less contrasted rings in the main disk. \citet{Block2006} proposed that the collision partner was a M-32 like galaxy, comprising about one-tenth of the mass (dark matter included) at the beginning. After stripping experienced in the collision, the M\,32 mass is now 1/23 that of the main target, M\,31. The M\,32-like plunging head-on with an impact parameter of 4~kpc, at a relative velocity of 265~\kms 210 Myr ago, has triggered the propagation of an annular wave, now identified with the 10~kpc ring; there is also a second wave propagating more slowly behind it \citep[e.g.,]{1996FCPh...16..111A}, which would correspond to the inner ring. In addition, the inner ring has formed and is propagating in a tilted and warped disk, which accounts for its almost face-on appearance, in contrast with the inclined main disk of M\,31 \citep[][]{1985ApJ...290..136J}. This scenario explains why the cold gas has been expelled from the central region and also why there are shocks and hot gas present in the inner kpc. We must note that in addition to the large inclination of the main disk \citep[e.g.,][]{2001ChPhL..18.1420M}, there are various shells connected to the large scale structures superposed on this two-ring morphology \citep[e.g.,][]{2022AJ....164...20E}, adding to the complexity.

A stellar bar, or more exactly a triaxial bulge, has long been detected in the center of M31 \cite[e.g.,][]{Stark1977, Athanassoula2006, Beaton2007}, in particular, due to the isophote twist. According to \citet{Saglia2018}, there is an axisymmetric classical bulge, with old age ($>$ 10 Gyr) and high metallicity in the M31 center, showing a negative [Z/H] gradient. No signature of the bar or boxy bulge is seen in the age or [$\alpha$/Fe] maps, which are approximately axisymmetric. However, in Fig. 14 of \citet{Saglia2018}, we can see that the metallicity of the old stellar population is enhanced along the bar and boxy bulge, as seen in other barred galaxies. \citet{Blana2017} showed that a triaxial bulge with zero pattern speed does not reproduce the measured stellar kinematics. A box-peanut component with a non-zero pattern speed must be the result of the buckling instability of a (previously thin) bar, although there is not presently any thin bar outside the boxy-bulge radius. The near-infrared isophotes outside this radius align onto the major axis of the galaxy. A more detailed discussion of a possible bar in the disk of M31 is reported in Sect.~\ref{sect:discuss}.

In addition to being a quiescent galaxy with little star formation, M31 has an ultra-weak nuclear activity \citep{2000ApJ...540..741D}.
\citet{2000MNRAS.312L..29M} first detected only a small amount of molecular gas ($1.5 \times 10^4 M_\odot$) within 1.3' (305~pc in projection). Relying on NOEMA interferometric observations, \citet{2019A&A...625A.148D} detected small molecular gas clumps within 250~pc, corresponding to $8.4\pm0.4 \times 10^4 M_\odot$ \citep[see also][]{2017A&A...607L...7M}. The molecular gas is very clumpy and no ongoing star formation has been detected in this central region \citep[e.g.,][]{2006AJ....132..271O,2009MNRAS.397..148L,2011AJ....142..139A}. In \citet{Melchior2011}, we detected molecular gas with IRAM-30m with very large line splittings up to 260\,km\,s$^{-1}$ in the northwest (NW) side of the disk along the minor axis. Towards some lines of sight, there were even three distinct velocity components. We discussed the fact that these apparently counter-rotating components appeared to be compatible with the \citet{Block2006} scenario, implying an inner ring with a tilted inner disk superposed on the main disk. In this framework, the observed narrow component was interpreted as the ring, the second wider peak was associated with the inner disc, and the third weak mid-wide component at the systemic velocity was thought to be associated with the main disk seen in projection.  In \citet{Melchior2016}, were presented new molecular observations of dense gas on both sides of the minor axis. This further supported the previous scenario, as the narrow and wide components were at opposite velocity on both sides of the major axis.
 
Some ionized gas has also been widely detected in the central field \citep[e.g.,][]{1971ApJ...170...25R,Ciardullo1988,1987A&A...178...91B,2008MNRAS.388...56B,2010MNRAS.404.1879L}. This ionized gas \citep[approximately 1500 M$_\odot$ according to][]{1985ApJ...290..136J} can be accounted for by mass lost from evolving stars. 
Recent works discussed a possible very weak star formation activity \citep{2022AJ....163..138L} compatible with the clumpy molecular gas distribution.

\begin{figure*}[h]
\centering 
\includegraphics[width=0.50\textwidth]{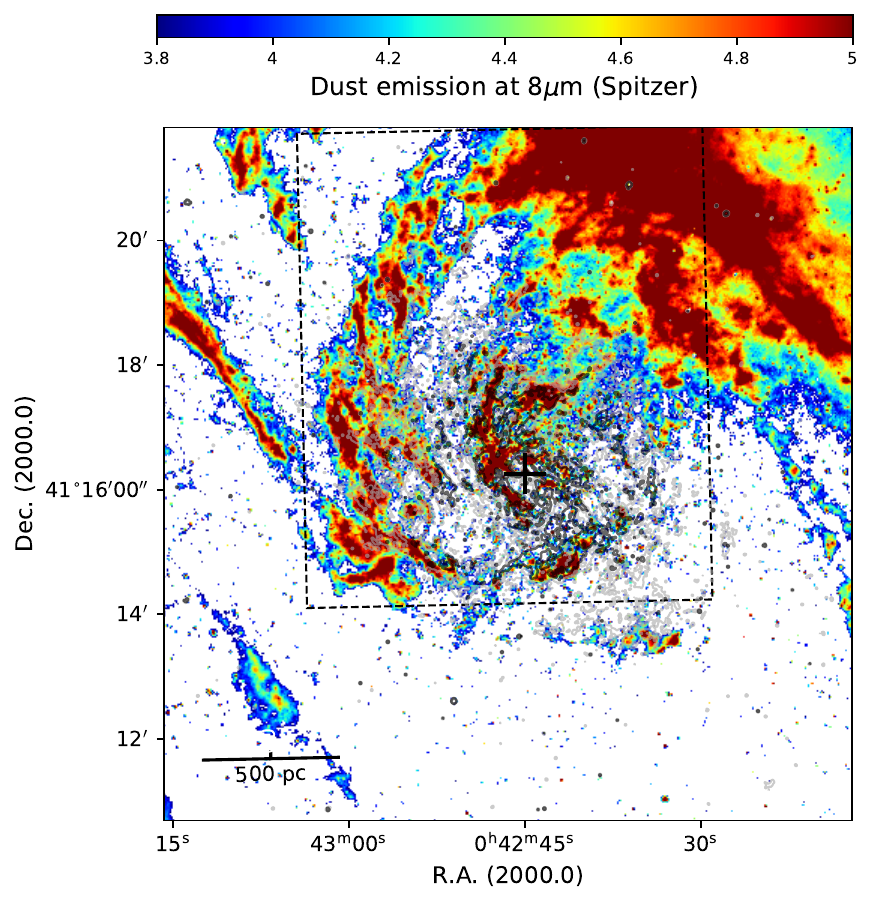}
\includegraphics[width=0.482\textwidth]{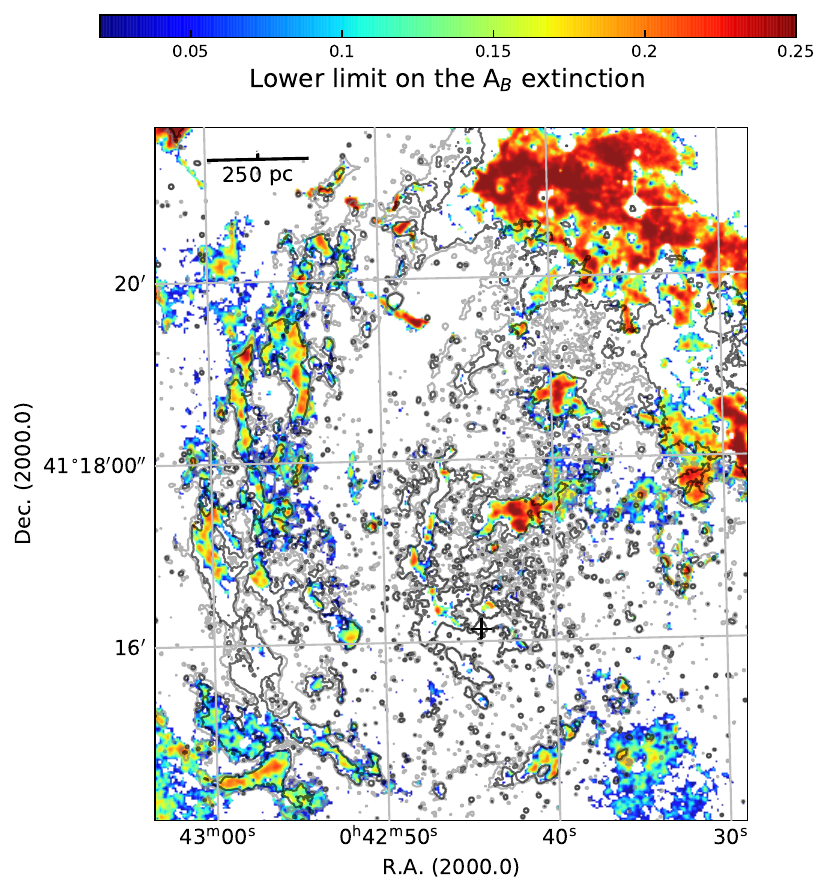}
\caption{Dust location in the central kpc field of view of Andromeda. Left panel: Dust emission map obtained from the 8$\mathrm{\mu}$m Spitzer map with a subtraction of the stellar continuum at 3.6$\mathrm{\mu}$m. \citep{Block2006}. The contours correspond to the ionized gas intensity displayed in the top left panel of Fig.~\ref{fig:kin}. The dashed square corresponds to the inner ring region displayed in the right panel. Right panel: an upper limit estimate on the $A_B$ extinction \citep{2000MNRAS.312L..29M}.
The gray contours correspond to the dust emission at 8\,$\mathrm{\mu}$m displayed in the left panel. The black cross symbol represents the M31 center.}
\label{fig:ext}%
\end{figure*}

In the present article, we further investigate this central kpc field, relying on different observations achieving some kinematic estimates; namely, optical ionized gas spectroscopy with the imaging Fourier transform spectrometer SITELLE at CFHT and molecular gas observations with the IRAM-30m telescope.
In a first step, we present a dynamical modeling based on tilted rings embedded in a global potential, mainly due to stars, as the gas mass is very weak in this region ($\sim$ 1500\,$M_\odot$). We relied on a kinematic map extracted from SITELLE observations with the SN3 filter targeting the H$\alpha$ and [NII] lines \citep{2018MNRAS.473.4130M}. In a second paper, we will present the full exploitation of the three SITELLE data cubes with a proper stellar continuum subtraction with Nburst \citep{2007MNRAS.376.1033C} and excitation diagrams. 

In the model, we follow the disk tearing methodology proposed by \citet{Raj2021} to account for misaligned disks spinning around galaxy nuclei. These authors showed that in the settling process to align angular momenta, thin and highly inclined precessing disks are more likely to yield large warp amplitudes, while low-viscosity disks are generally expected to become unstable and break in one or two rings. Although the origin of the torques are different, this scheme is quite adapted to the characteristics of the gas in the central region of the highly inclined disk of M31. In Sect.~\ref{sect:obs}, we describe the data we obtained with SITELLE at CFHT, and we extract in Sect.~\ref{sect:proc} a perturbed kinematic map derived from the optical ionized gas ([NII], H$\alpha$, [SII]). In Sect.~\ref{sect:dyn}, we present the characteristics of the modeling that we developed to reproduces the velocity field. The main features can be well reproduced with a customized static modeling of a central tilted disk extending to a warp region connecting the main disk and an independent offset 1-kpc ring. In Sect.~\ref{sect:fitting}, we discuss the main results achieved, that is how the proposed modeling reproduce the data. In Sect.~\ref{sect:discuss}, we compare our results to previous works, in particular, focusing on the proposition that these features correspond to a bar. Our conclusions are presented in Sect.~\ref{sect:conclu}. Throughout this paper, we adopt a distance to M31 of 780~kpc, thus, 1~arcsec~=~3.8~pc.

\section{Data}
\label{sect:obs}
Here, we present the observational data we used to constrain the dynamical modeling. 
In Sect.~\ref{ssect:co}, we discuss the archival data from molecular gas observations at IRAM-30m. 
In Sect.~\ref{ssect:sitelle}, we recall the characteristics of the SITELLE data cube used to extract the kinematics. In Sect.~\ref{ssect:dust}, we present the 2D maps from archival data, namely, the dust emission and upper limits on dust extinction.
% mapping the central region in the infrared and in optical extinction, that can be compared with the ionized gas detection positions. We also describe the two sets of new kinematic observations, in ionized and molecular gas. The characteristics of these data will be subsequently used to constrain the proposed modeling.

\subsection{Archival molecular gas data}
\label{ssect:co}

To make a comparison with the molecular gas component, we again reduced the CO(2-1) data from IRAM-30m, taken with the HEterodyne Receiver Array (HERA) focal plane array, between November 2011 and March 2012 \citep{Melchior2013}. The spatial resolution is 11'' and spectral resolution 2.6~\kms, we refer to the above paper for more details. We have built a data cube, with pixel size of 5.33'', and 13~km/s. The field of view of the cube is 7.2 $\times$ 7.9$'$, corresponding to 1.63 $\times$ 1.79~kpc.

\subsection{SITELLE SN3 data cube}
\label{ssect:sitelle}
We rely on a data cube obtained with SITELLE and installed at CFHT\footnote{The data cube is available on the CADC archives and can be downloaded on: 
\href{https://ws.cadc-ccda.hia-iha.nrc-cnrc.gc.ca/raven/files/cadc:CFHT/1982678p.fits}{https://ws.cadc-ccda.hia-iha.nrc-cnrc.gc.ca/raven/files/cadc:CFHT/1982678p.fits}}.
The observations were performed on August 24, 2016, with the SN3 filter designed to detect the H$\alpha$-6563$\AA$ Balmer line and the [NII]-6548,6583$~\AA$ and [SII]-6716,6731 doublets. The $11^\prime \times 11^\prime$ field of view has been centered on M31 optical nucleus \citep[J2000: RA: 00h42m44.37s, DEC: $41^\circ 16^\prime08.34^{\prime\prime}$;][]{1992ApJ...390L...9C}. This field has been integrated for 4.1 hours with a pixel size of 0.321$^{\prime\prime}$. Such an interferometric technique enables us to reach a good spectral resolution (here $R=4800)$, but it suffers from the multiplexing disadvantage \citep{2013ExA....35..527M}: all the photons collected by each interferogram are redistributed over all the channels of each spectrum. This effect is important for the central region of M31, but has been limited with the use of the SN3 filter. The reduction of the data cube has been described in \citet{2018MNRAS.473.4130M}. This region devoid of star-forming regions is dominated by diffuse ionised gas (DIG), as discussed in Sect.~\ref{sect:proc}.

\subsection{Archival dust extinction and emission maps}
\label{ssect:dust}
As displayed in Fig.~\ref{fig:ext}, the central kpc field of view of Andromeda has been mapped in dust emission by \citet{Block2006}. The map in the left corresponds to the 8\,$\mathrm{\mu}$m Spitzer map with the stellar continuum at 3.6\,$\mathrm{\mu}$m \citep{Block2006} has been subtracted. It has been overlaid with contours corresponding to the H$\alpha$ map (discussed in the next section). On the right panel, the $A_B$ extinction map, computed assuming the dust is seen in front of the bulge, is presented, as discussed in \citet{2000MNRAS.312L..29M}. The apparent extinction is larger on the near part of the disk (NW), which explains why the H$\alpha$ emission is weaker in the northwest region. 

Both maps show the presence of an off-center inner ring at 1~kpc. These maps provide some constraints on the geometry of this ring detected in dust emission in the northwest part, while it is emitting in ionized gas more strongly in the southeast (SE). The overall dust and gas content in this region is weak, and we went on to assume that the gas is dominated by the stellar potential.
In this paper, we further explore the modeling of this region with the support of the new SITELLE and new CO(2-1) velocity maps of gas observed along the minor axis, to test this scenario. 
%Figure~\ref{fig:complex2} highlights several structures on the ionized-gas velocity map that will be used as constraints in the proposed modeling. The label 1 corresponds to a region representative of the main M31 disk. The label 2 points to the location of the ring as detected in the dust emission, and for which we detect a distinct velocity in the ionized gas. The label 3 highlights regions with high velocity gradient. The label 4 exhibits a small inner disk, probably titled with respect to the main disk, identified by a kinematic twist signature.

\subsection{Off-centering of the multiple components}
\label{ssect:off}
As discussed in \citet{Block2006}, the inner and 10~kpc rings are off-centered. The galaxy itself is clearly perturbed and non symmetric. 
However, while the central velocity field looks a priori asymmetric (Fig.~\ref{fig:kin}), the central circumnuclear disk is only slightly off-centered (Fig.~\ref{fig:center}).
We assume that the disk and warp components are centered, but we do introduce extinction, depending on the depth of each component. As later discussed, this will introduce some asymmetry. 

In parallel, the inner ring displayed in Fig.~\ref{fig:ext} is clearly off-centered, which makes it an improbable inner Lindblad resonant ring, in the case of a bar scenario. We rely on its apparent position in the Spitzer map (center, position angle, and axis-ratio) to constrain its position.
\begin{figure}[h]
\centering
  \includegraphics[clip,width=\linewidth]{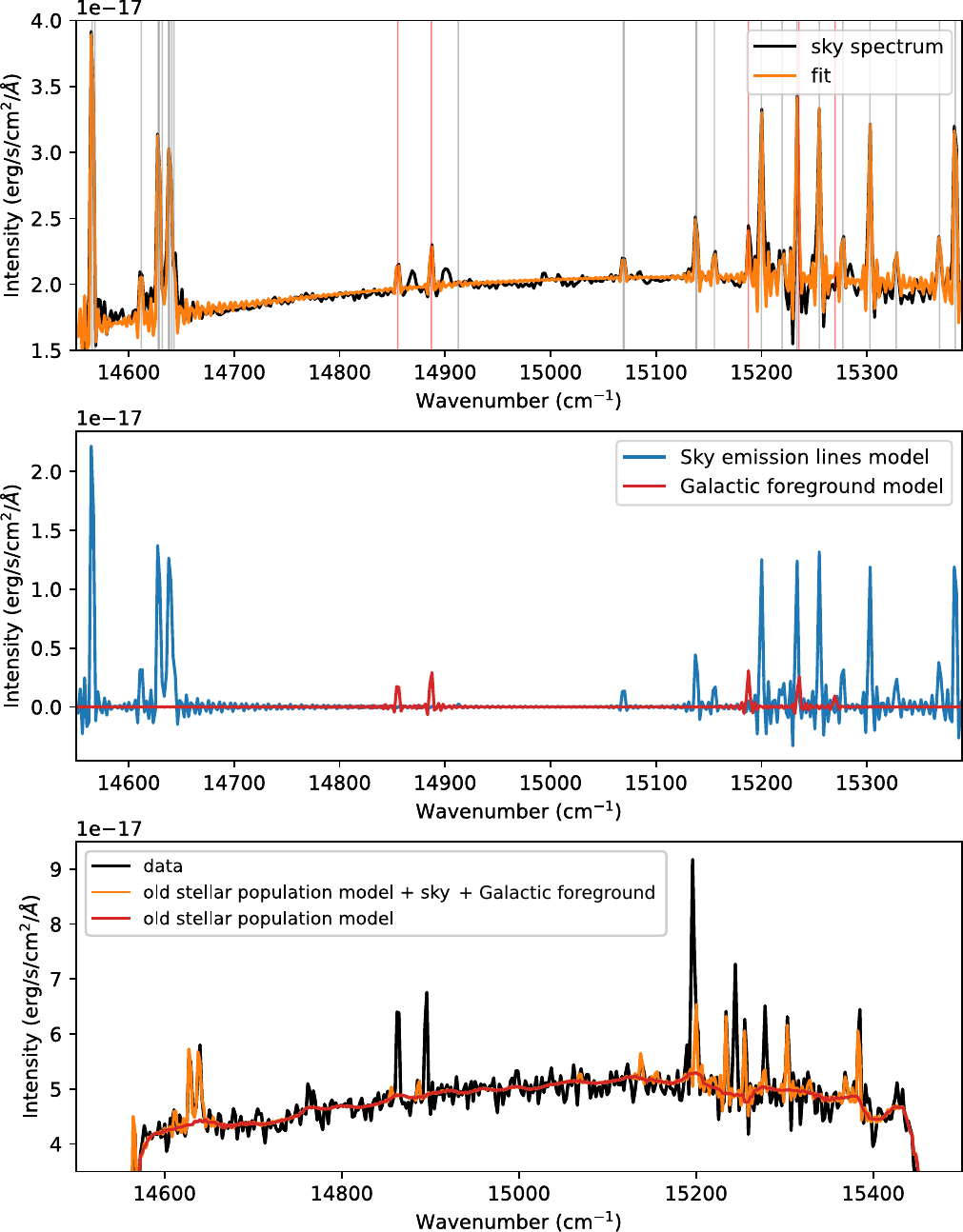}
  \caption{Background subtraction. Top panel: Fit of the sky lines over a low brightness region of the field of view. The vertical gray lines indicate the position of the atmospheric sky lines. The five red vertical lines indicate the position of the foreground emission of the DIG of our own Galaxy in the $[NII]$(6563,6584), $[SII]$ $(6717,6731)$ and H$\alpha$ lines at a velocity of -38$\pm$5\,km/s.
  Middle panel: Sky lines model and Galactic foreground model. During the fitting procedure the velocity of the sky lines and the velocity of the Galactic foreground were two independent parameters. 
  Bottom panel: Analyzed spectrum and model of the background.}
    \label{fig:background_modeling}
\end{figure}

\section{Extraction of the kinematics from the SITELLE SN3 data cube}
\label{sect:proc}
While a complete analysis of the data cubes is in progress, we rely on a kinematic map extracted from the data for the purposes of this study, as follows. The idea is to optimize the extraction of the kinematic information and the small pixel size. Indeed, the brightness of the DIG is low when compared to the surface brightness of the old stellar population background in the center of M31. Even if its presence is detectable in some frames of the spectral cube, we must start by a careful removal of the other components of the background spectrum in order to reveal the full content of its emission. We can then detect its different components and estimate their flux as well as their velocity which can finally be used as an input for a fitting procedure. 
We describe these three steps with more details in the following.

In Sect.~\ref{ssect:bg}, we discuss how we modeled the stellar background. In Sect.~\ref{ssect:detect}, we discuss the automatic detection of the gas velocity relying on the five emission lines. In Sect.~\ref{ssect:linefit}, we present the line fitting used to get the velocity dispersion map.

\subsection{Background modeling: Sky lines, Galactic foreground, and stellar continuum}
\label{ssect:bg}
This background spectrum is indeed the superposition of the old stellar population continuum and the atmospheric OH lines, neither of which are trivial to estimate.
As the bulge stellar emission is strong in the whole field of view, there is no region where the atmospheric emission can be directly estimated. We could nevertheless find two corners, northeast (NE) and southwest (SW), where the bulge emission (and especially its H$\alpha$ \,absorption line) is small enough that its effect on the emission lines can be partially neglected. We combined these regions to obtain a spectrum which was fitted to compute a model of the foreground emission. This emission contains both the atmospheric sky lines and some Galactic emission in $[NII]$(6563,6584), $[SII]$ $(6717,6731)$ and H$\alpha$ at a mean velocity of -38$\pm$5\,km/s. (see Fig.~\ref{fig:background_modeling}). The Galactic emission seems homogeneous both in terms of flux and velocity in the whole field of view so that once subtracted, it should have a negligible impact on our estimation of the emission of the DIG of Andromeda.

\begin{figure}[h]
\centering
  \includegraphics[clip,width=\linewidth]{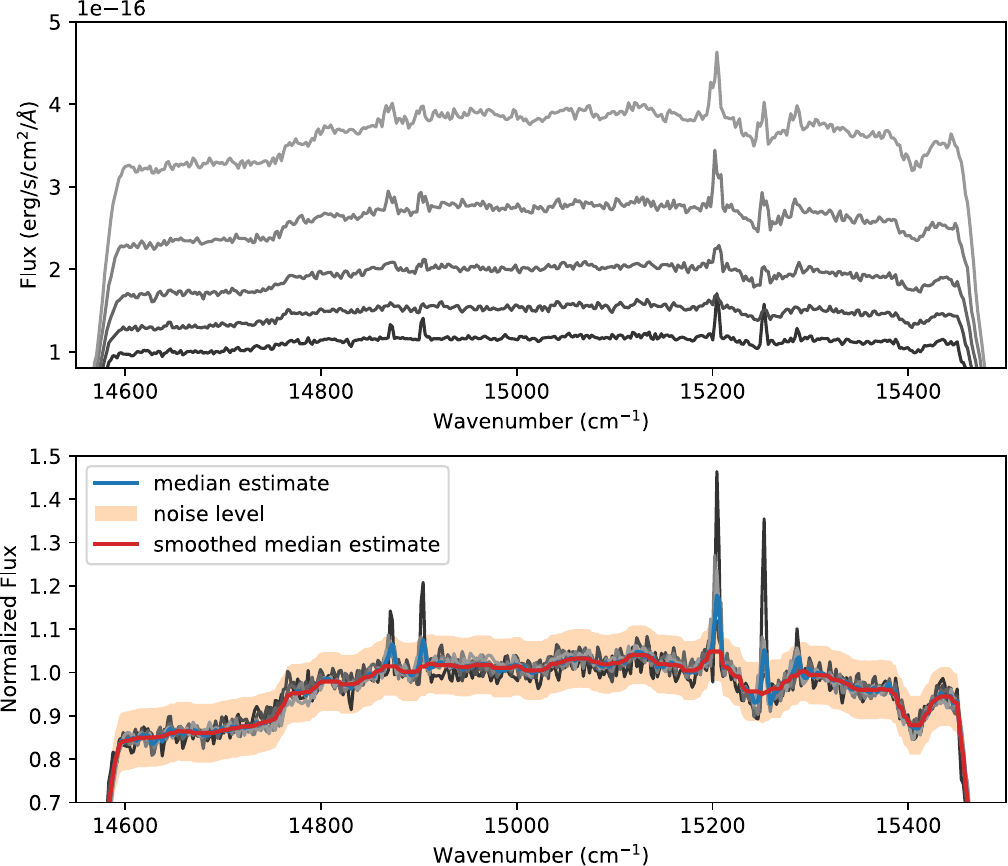}
  \caption{Example of the old stellar population modeling in a 200$\times$200 pixels$^2$ box near the center of the galaxy. Top panel: Random spectra taken in the box to illustrate the strong brightness gradient. Bottom panel: Same spectra after a normalization by their mean flux. The resulting median spectrum of all the normalized spectra contained in the box is shown in blue, the smoothed version used as an estimation of the old stellar population is shown in red. The noise level of the individual 7$\times$7 binned spectra used during the fitting procedure is shown as an orange envelope.}
    \label{fig:background_modeling3}
\end{figure}

The old stellar population emission varies considerably in the field of view but at a sufficiently large scale that it can be approximated over large bins of data. We choose a binning surface of 200$\times$200\,pixels$^2$ because it is small enough to capture the variations of the old stellar population emission and large enough that the DIG emission varies sufficiently in velocity that a median spectrum taken over this region is not significantly contaminated by it (see Fig.~\ref{fig:background_modeling}). Another challenge comes from the fact that the emission level displays a strong pixel-to-pixel gradient, especially near the center of the galaxy. As the emission (and particularly the absorption lines) can be approximately considered to be proportional to the mean level of emission in the observed band, a naive mean spectrum of a large region would be dominated by the spectra of the brightest pixels. In order to get a more balanced estimation of the spectrum covering a binned region, the spectra were normalized by their mean emission level before being median-combined resulting in a normalized estimation of the background spectrum of the binned region. This last spectrum was then smoothed to remove the remaining emission-lines and obtain a normalized model of the galactic background in the 200$\times$200\,pixels$^2$ box (see Fig.~\ref{fig:background_modeling3}). Using an old stellar population modeling instead of a simple smoothing is certainly a more precise approach, but the noise level of the individual spectra used for fitting (even binned 7$\times$7) is high enough that we estimate the difference between both processes (smoothing vs stellar population modeling) to be smaller than a few percent and thus negligible. The normalized model of the galactic background at a particular position is then linearly interpolated between the positions of the binned spectra and multiplied by the mean emission at the same position (see Fig.~\ref{fig:background_modeling}). 

\begin{figure}[h]
\centering
  \includegraphics[clip,width=\linewidth]{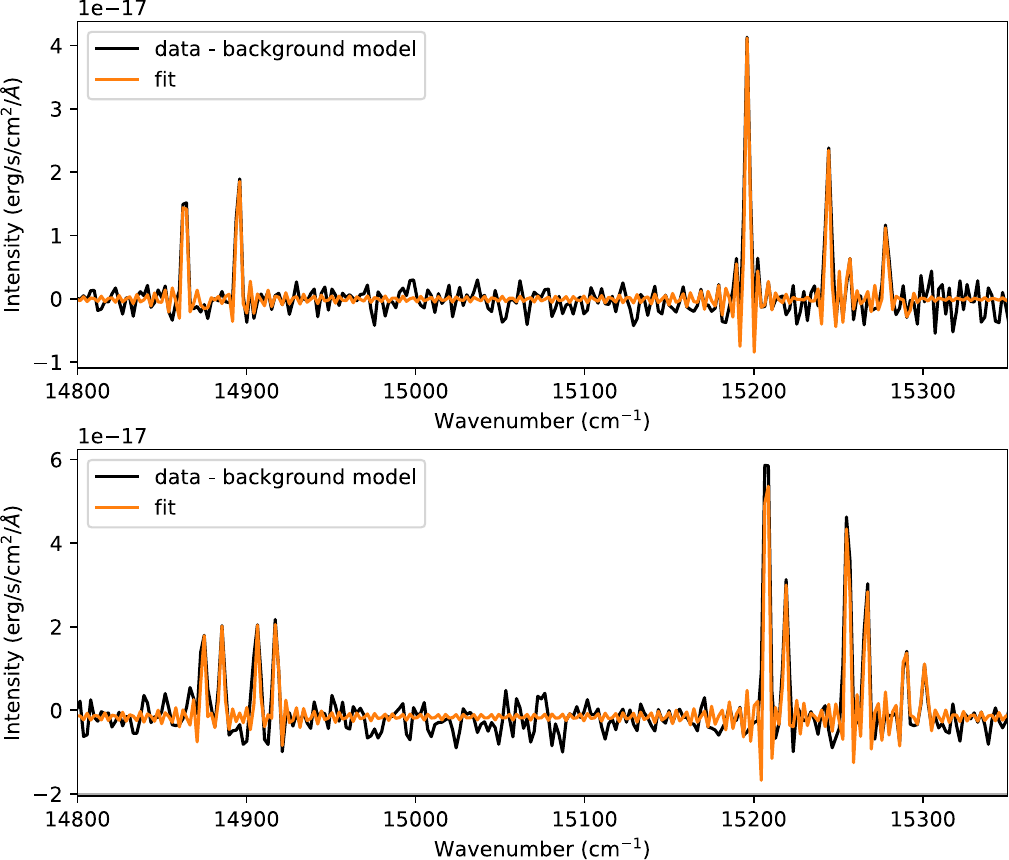}
  \caption{Examples of emission lines fitting on background corrected spectra displaying only one (top panel) or two components (bottom panel). Top panel: Fit over a background subtracted spectrum revealing the pure emission of the DIG. Bottom panel: Spectrum displaying two resolved velocity components along the line of sight. }
    \label{fig:background_modeling2}
\end{figure}

\begin{figure*}[h]
\centering 
\includegraphics[width=0.48\textwidth]{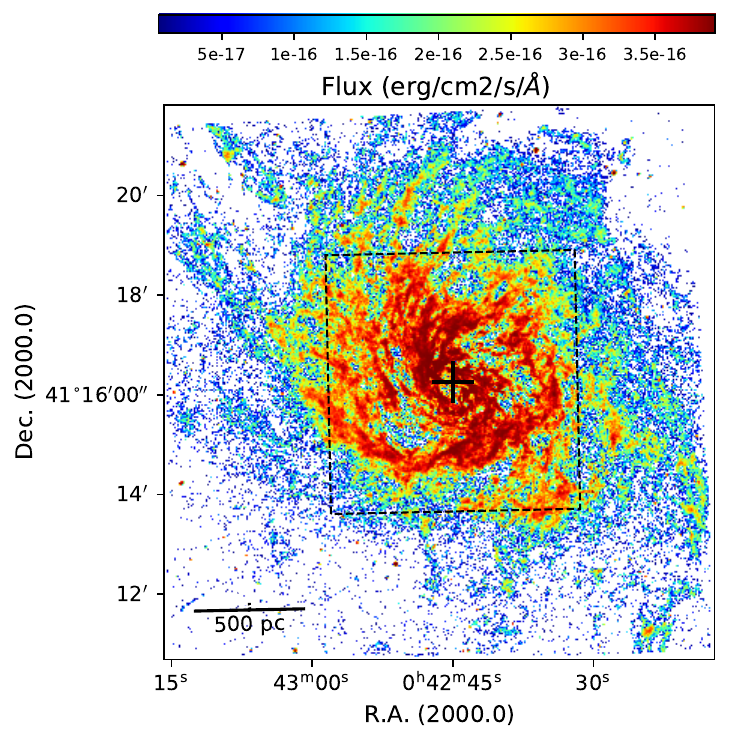}
\includegraphics[width=0.48\textwidth]{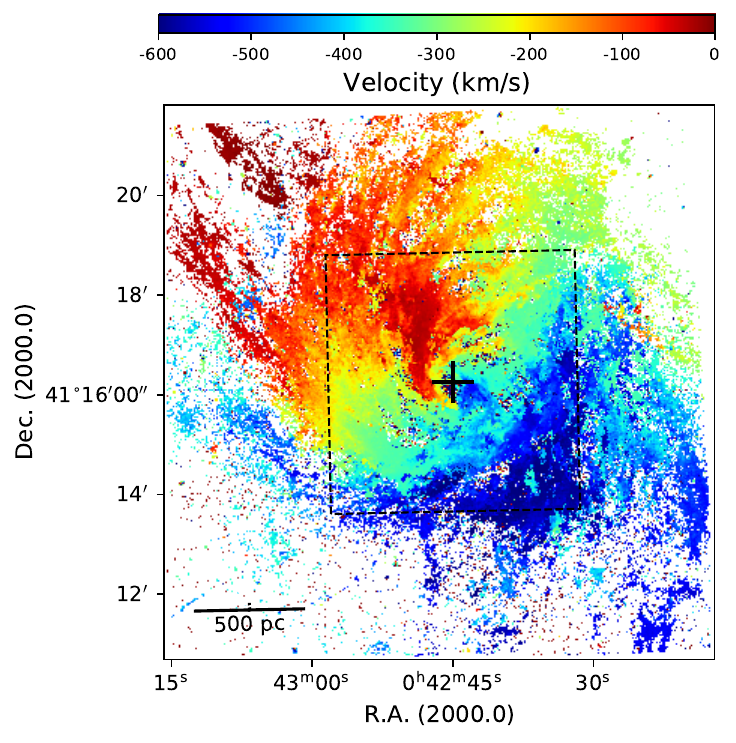}
\includegraphics[width=0.48\textwidth]{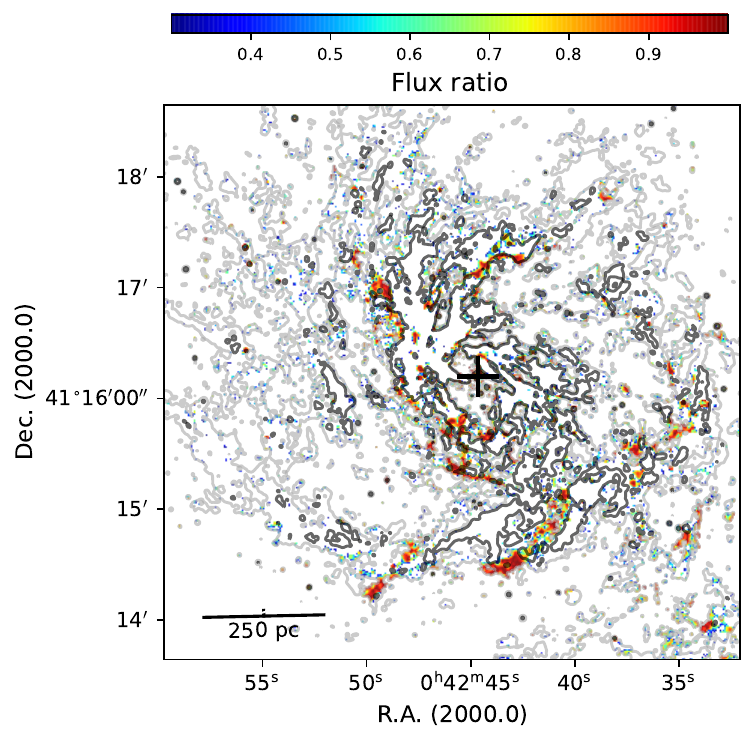}
\includegraphics[width=0.48\textwidth]{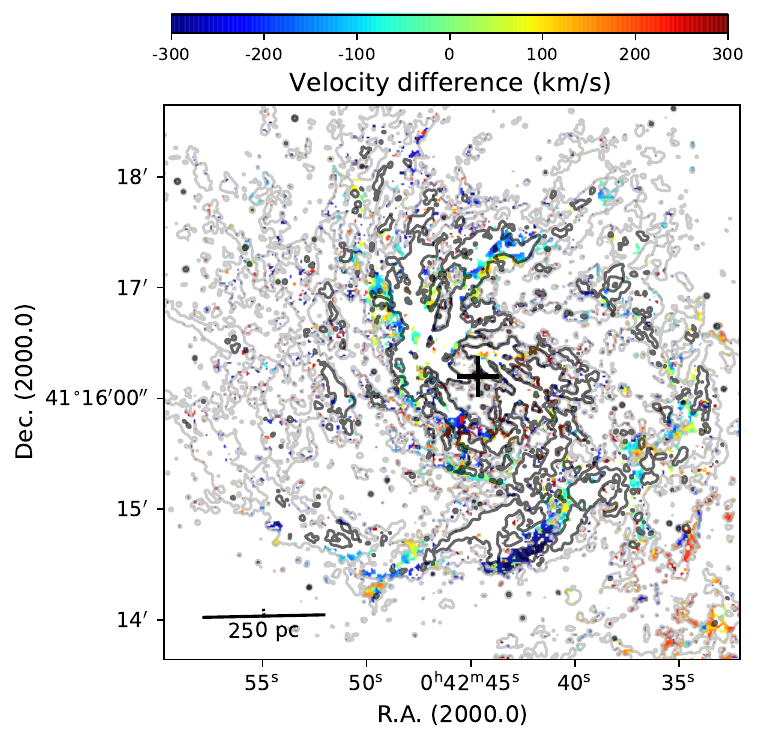}
\caption{Intensity and kinematic information extracted from the ionized gas with SITELLE SN3 data cube. Top panels: Ionized gas intensity (left) and the heliocentric velocity (right) corresponding to the main velocity component. The ionized gas intensity has a threshold of $6\times 10^{-18}$\,erg\,cm$^{-2}$\,s$^{-1}$\,$\AA^{-1}$. The dashed squares correspond to the field of view displayed in the bottom panels. Bottom panels: Flux ratio $\phi_2/\phi_1$ (left) and velocity difference $V_2 - V_1$ (right) for the second kinematic component with respect to the first one. The contours correspond to the intensity levels ($1.8\times 10^{-17}$, $3.6 \times 10^{-17}$\,erg\,cm$^{-2}$\,s$^{-1}$\,$\AA^{-1}$) of the flux map of the main component (top left). The black cross symbol representing the M31 optical center, 00h42m44.37s +41d16m08.34 \citep{1992ApJ...390L...9C}.}
\label{fig:kin}
\end{figure*}

\subsection{Automatic detection of the emission components of the DIG}
\label{ssect:detect}
The automatic detection of the different emission components of the DIG was achieved with the same algorithm as the one used in \citep{2021MNRAS.502.1864M}. This algorithm is based on the correlation of the spectra with a five emission lines comb ($[NII]$(6563,6584), $[SII]$ $(6717,6731)$, and H$\alpha$). We used it on $7\times 7$ binned spectra and we obtained a map of the estimated velocity of the DIG at each binned position.
In some cases, two velocity components are found along the same line of sight. As discussed in the next section, an example of such two components is displayed in the bottom panel of Fig.~\ref{fig:background_modeling2}. The derived velocity map is displayed in the top right panel of Fig.~\ref{fig:kin}, as well as the corresponding intensity (top left panel). 
Error bars on these velocities are estimated between 10 and 60~\kms, according to the flux signal-to-noise ratio.
When two velocity components can be detected, we display their flux ratio, and velocity difference in the bottom panel of Fig.~\ref{fig:kin}. 

\begin{figure}[h]
   \centering
   \includegraphics[width = 0.49\textwidth]{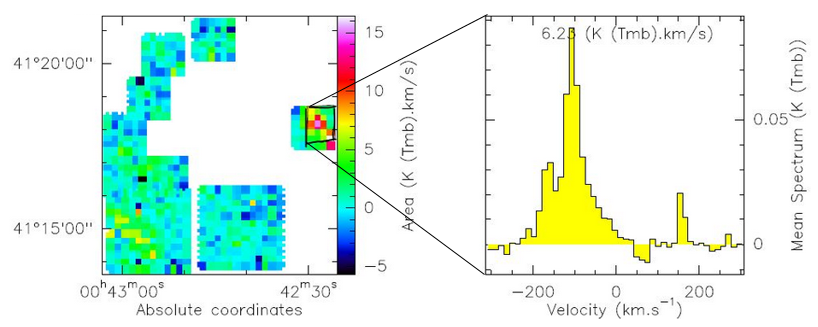}
    \includegraphics[width = 0.49\textwidth]{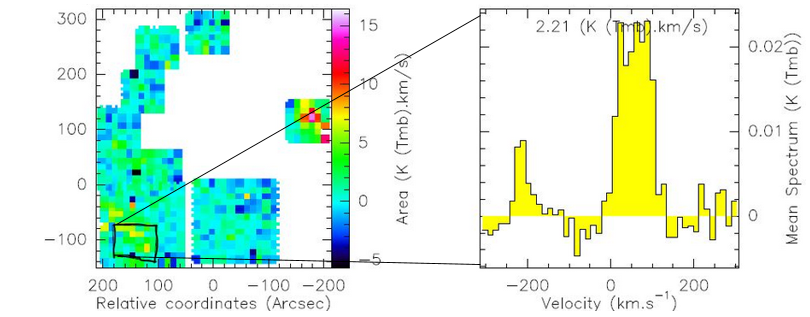}
   \caption{CO(2-1) spectra averaged over $\sim$ 60'' regions in diameter in the northwest (top) and the southeast (bottom) of the observed IRAM-30m moment-zero map. The NW corresponds to (-160'',120'') offsets, and the SE to (150, -110'') offsets relative to the M31 center RA=00:42:44.35 DEC=41:16:08.6. They correspond to the same regions as Fig.~\ref{fig:Ha-spec} }
   \label{fig:NWSE0}
\end{figure}

\begin{figure}[h]
   \centering
   \includegraphics[width = 0.4\textwidth]{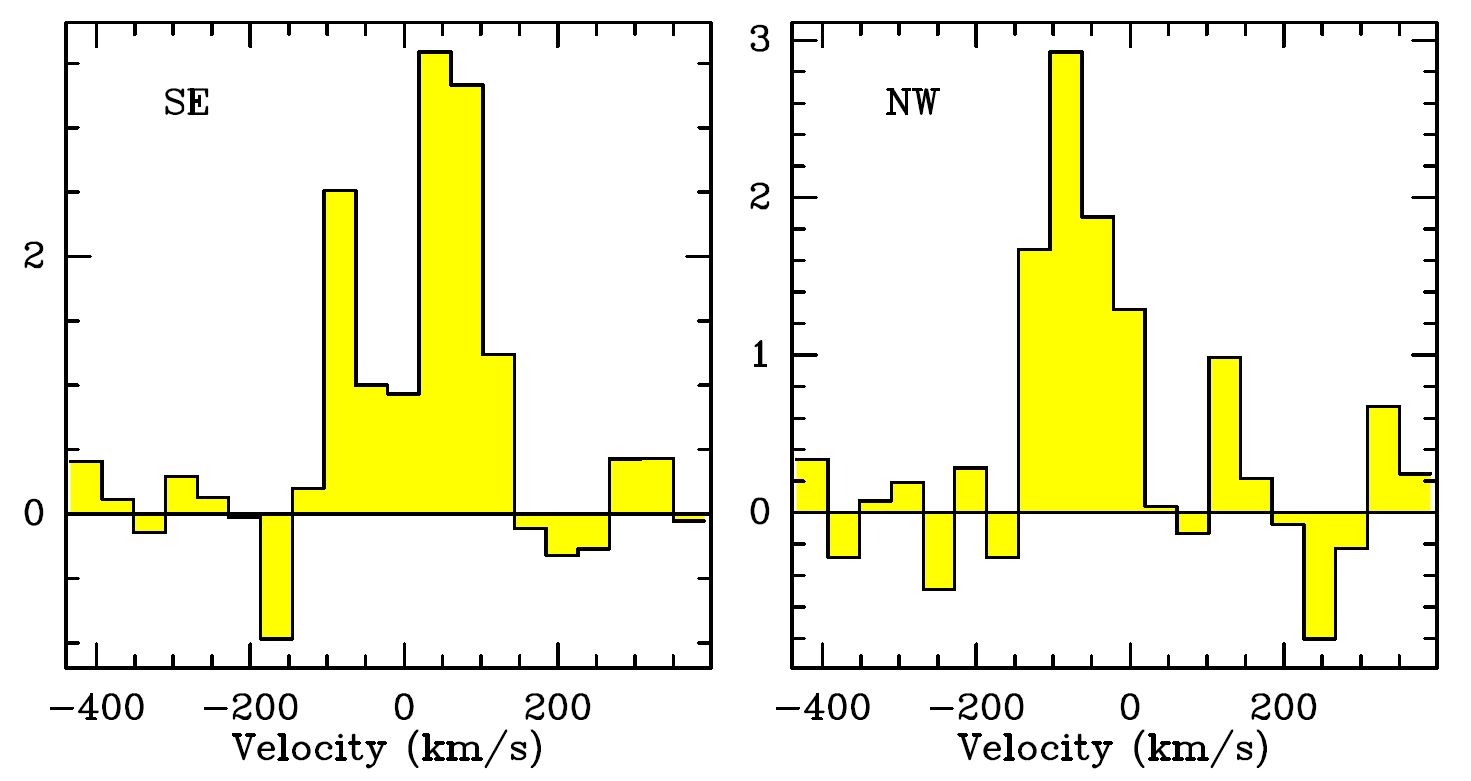}
   \caption{H$\alpha$ spectra averaged over 12'' regions in the southeast (left) and the northwest (right) of the observed SITELLE map. The SE corresponds to (124'',-72'') offsets, and the NW to (-199, 72'') offsets relative to the M31 center RA=00:42:44.35 DEC=41:16:08.6.}
   \label{fig:Ha-spec}
\end{figure}

In parallel, the most remarkable features of the molecular gas observations are presented in Fig.~\ref{fig:NWSE0}. Two components are detected along the minor axis, where a single disk would be expected with a systemic velocity. In addition, one of the two has a large velocity dispersion, as expected for a large velocity gradient (typical of an inclined disk in a 12-arcsec beam). The second one is relatively weak and narrow. In \citet{Melchior2011,Melchior2013}, we proposed that this narrow component could be accounted for by an offset ring as revealed by \citet{Block2006}. On the northwest side, the broad (resp. narrow) component is blue-shifted (resp. red-shifted), while on the southeast side, it is the opposite. 

\begin{figure}[h]
\centering 
\includegraphics[width=0.45\textwidth]{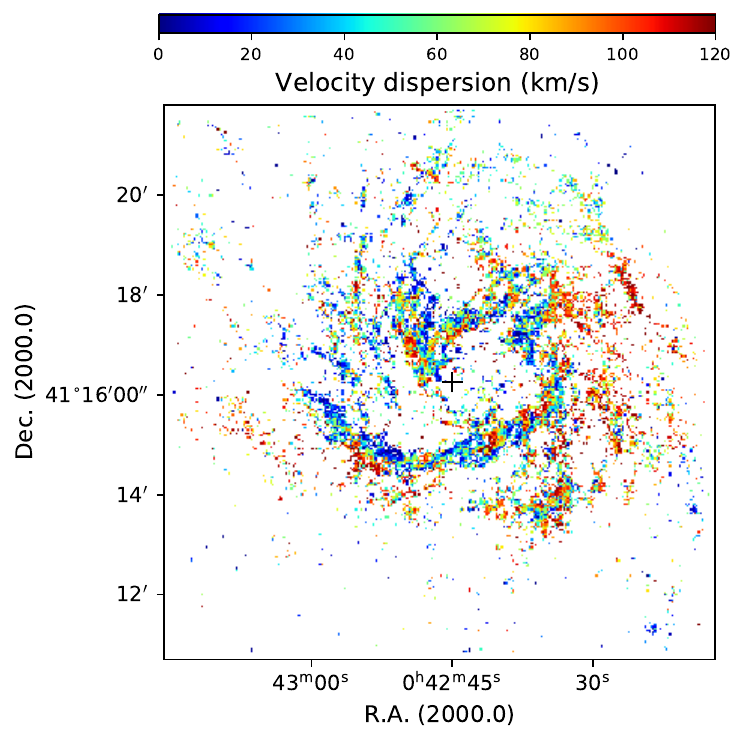}
\caption{Velocity dispersion derived from the Gaussian fit on SITELLE sky-subtracted cube.}
\label{fig:sig}
\end{figure}

\subsection{Line fit to extract the velocity dispersion}
\label{ssect:linefit}
\begin{figure}[h]
\centering
  \includegraphics[clip,width=0.8\linewidth]{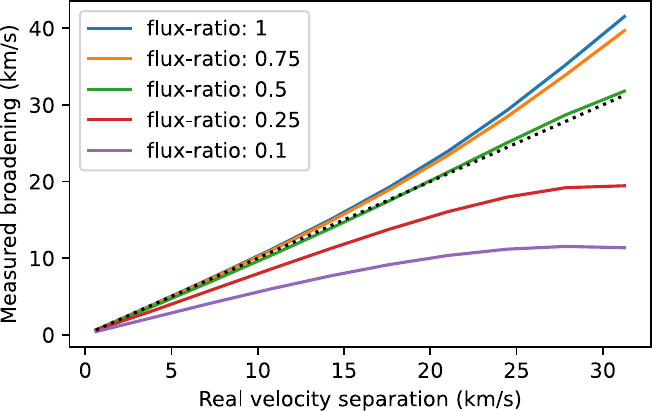}
  \caption{Measured broadening when using a \texttt{sincgauss} model (broadened Gaussian emission-line) to fit two unresolved emission-lines. We have simulated the fit with a \texttt{sincgauss} model of two unresolved emission-lines with different flux ratios. We can clearly see that the measured broadening can be very different from the real velocity separation and depends strongly on the flux ratio of the unresolved lines. Note that, at the resolving power of our data ($R=4800$), two components are considered resolved when separated by more than 63~\kms. The dotted line is the bisectrix, meaning equality between the two quantities.}
    \label{fig:2sinc}
\end{figure}

In addition to the positions where two ionized-gas components are clearly resolved along the line of sight and with comparable intensity (see bottom panel of Fig.~\ref{fig:background_modeling2}), in most cases, the (possibly multiple) components could not be spectrally resolved, but the lines were broadened. We used a sincgauss model \citep{2016MNRAS.463.4223M} to fit these lines and map their broadening (see Fig.~\ref{fig:sig}). Note that this model works under the hypothesis that the observed emission-line is the convolution of a Gaussian (the real emission-line) and an instrumental line function (a sinc). In some cases, given the spectral resolution of our data, this hypothesis might be reasonable, but, when the real emission-line is the sum of two different velocity components, the measured broadening of the Gaussian model is a biased measure of the velocity separation of the emission-lines. Depending on the flux ratio of the two unresolved emission-lines, the measured broadening may be off by more than 30\% from the real velocity separation (see Fig.~\ref{fig:2sinc}). The measured broadening must thus be considered as an indication of the existence of two or more unresolved components along the line of sight giving only an approximate insight on their real velocity separation. It might however be useful when the S/R is low enough to prevent the modeling of two different components with unknown fluxes -- which is the case in the present study.

%\vspace{10cm}
\subsection{Velocity dispersion and double-peaked emission profiles}
The previous kinematic map displayed in the right panel of Fig.~\ref{fig:kin} reveals a complex velocity field with discontinuities, which is capable of producing double-peaked features when smoothed in a larger beam. As displayed in Fig.~\ref{fig:Ha-spec}, spectra integrated with a 12-arcsec beam on the SITELLE-SN3 data cube display double-peak features at positions along the minor axis and on the east side. \citet{Opitsch2018} also mapped two components in the velocity field. In CO observations along the minor axis \citep{Melchior2011,Melchior2013}, spectra exhibits also two components, that we interpreted as an inner ring (corresponding to the thin component) and a disk (corresponding to the thick component). 

We also extracted the velocity dispersion from a single Sinc-Gaussian fit, as displayed in Fig.~\ref{fig:sig}. For this figure, a threshold in flux has been required, with more than $3\sigma$ significance. The sensitivity is low in the very center due to the multiplex disadvantage.
Interestingly, we can note that several regions exhibit a velocity dispersion larger than 100~\kms, in particular in the southwest region. However, taking into account the multiphase properties of this diffuse gas is beyond the goal of this paper. 

We also observe, in the bottom panels of Fig.~\ref{fig:kin}, that some positions present double peaks at the nominal resolution (1~arcsec). Interestingly, they lie at the edge of the densest regions, suggesting a warping, or a large inclination gradient along the line of sight. In the southern region, the velocity difference reaches 300\,km\,s$^{-1}$ and seems typical of the southwest region where velocities range between -400 and -650\,km\,s$^{-1}$, with patchy regions at different velocities.

\begin{figure}[h]
    \centering
    \includegraphics[width=0.4\textwidth]{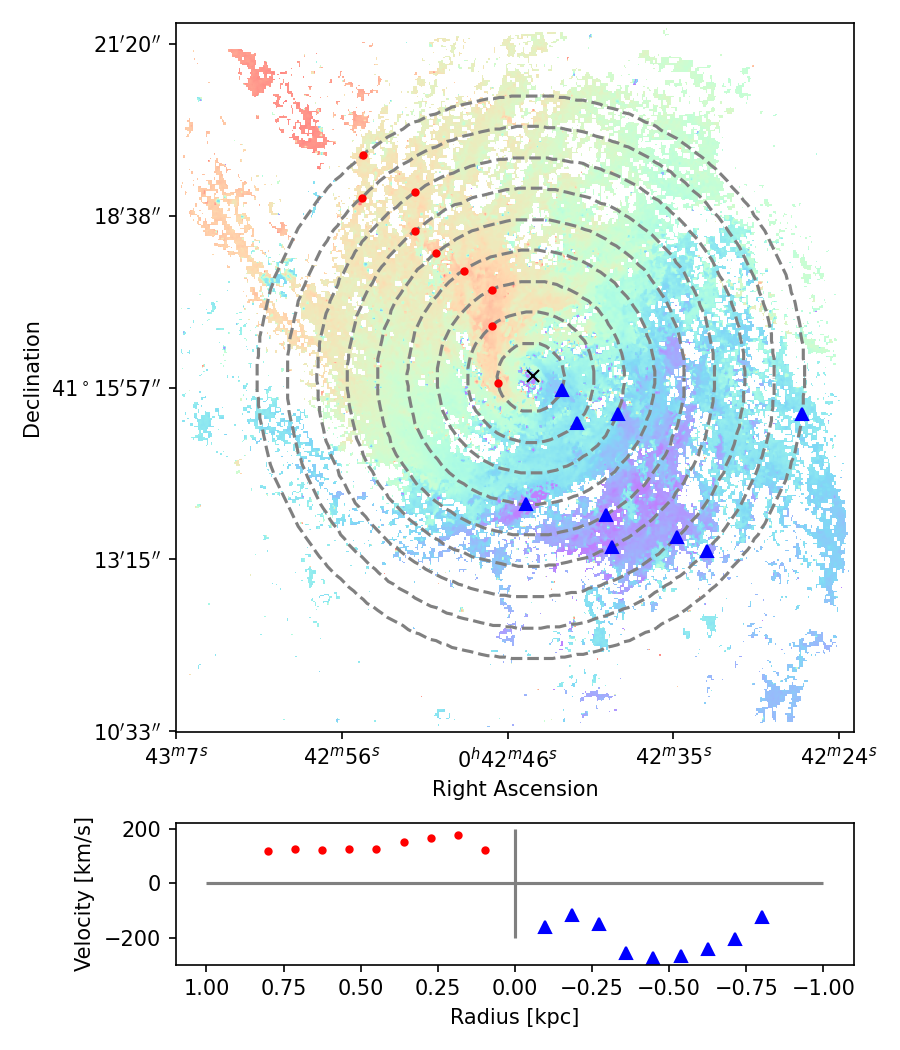}
    \caption{Simplified illustration of the map processing, on $N=9$ concentric circles, with radii regularly spaced between $R_{min}=0$ and $R_{max}=0.8$~kpc. 
    Top panel: Velocity map overlaid by the concentric circles (in gray dotted lines). The red dots (resp. blue triangles) are the maximal (resp. minimal) values $V_{max}$ (resp. $V_{min}$) for each circle, and have been computed with $n=7$.
    Bottom panel: Previous extrema $V_{max}$ and $V_{min}$ of the velocity displayed as a function of the distance from the center (i.e., the radius of its corresponding circle). The negative distances corresponds to the blue shifted part of the map, where the velocities are negative. The gray lines represent the origin axes.}
    \label{fig:minmax_process}
\end{figure}

\begin{figure*}[h]
    \centering
     \includegraphics[width=0.8\textwidth]{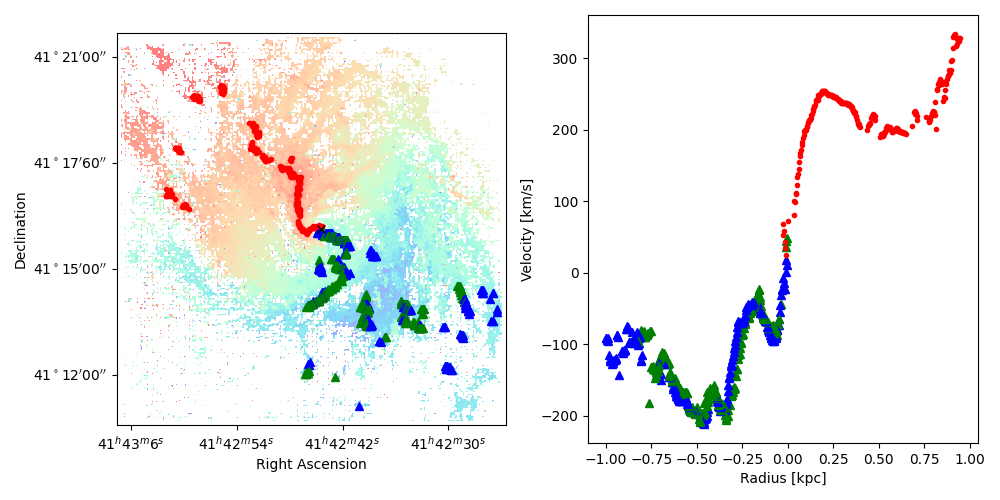}
    \caption{Minimal and maximal values, $V_{min}$, $V^{2nd}_{min}$, and $V_{max}$, of the velocity on $N=300$ concentric circles on the map, centered on the BH, with radii regularly spaced from $R_{min}=0$ to $R_{max}=1$\, kpc. 
    Left panel: Velocity map overlaid by the extreme points of the velocity. The red dots (resp. blue triangles) are the maximal (resp. minimal) values at each radii. As discussed in the text, the green points correspond to the $V^{2nd}_{min}$ velocity found when considering the second component detected and presented in Fig.~\ref{fig:kin}.
    Right panel: Minimal and maximal values of the velocity, as a function of the distance from the BH. The negative distances corresponds to the blue shifted part of the map.}
    \label{fig:minmax_all}
\end{figure*}

\subsection{Velocity map analysis}
To extract geometrical constraints from the velocity map, we searched the maximal and minimal values of the velocity at different radii. We were then able to identify regions with possibly independent morphology, and to derive their respective position angle (PA$=\theta$). In this Sect.~\ref{subsec:methodology}, we present the methodology used to extract information from the map. In Sect.~\ref{subsec:regions}, we subsequently identify the main regions of interest (or main patterns of the velocity field.).

\subsubsection{Methodology of the map analysis} \label{subsec:methodology}

We first computed a set of $N$ concentric circles of regularly spaced radii, expanding from the center of M31. The radius $r_i$ (in kpc) of each circle is defined as $r_i = i \times \frac{R_{max}-R_{min}}{N}$. On each of these circles, we selected the minimal and maximal values $V_{min}$ and $V_{max}$ of the velocity field displayed in Fig.~\ref{fig:kin}. Figure~\ref{fig:minmax_process} illustrates this process for $N=9$ circles.
Each circle $i$ with a radius $r_i$ (in kpc) is separated into $N_i=\frac{r_i}{10~kpc}\times 360$ arcs. Each arc then has a constant length of $\pi/180$~kpc corresponding to 17~pc.

We then defined a running box of $2 n \times 2 n$ pixels centered on each arc. For each box, we define an averaged velocity $V_{ave}=\frac{1}{4 n^2}\Sigma_{i=0}^{i=2n}\Sigma_{j=0}^{j=2n} V_{ij}$, with $n=7$. In parallel, we eliminated pixels with a value is smaller than $10\,\%$ of the maximum value in a similar box with $n=10$. Hence, for each circle, we kept the following boxes:
\begin{itemize}
    \item Pixels with a defined flux (i.e., above a threshold that depends on its surroundings), and boxes with less than 50\,$\%$ undefined-flux pixels;
    %\item The velocity dispersion, hence the uncertainty on the estimate value of the velocity, is low enough.
    \item The standard deviation of the velocity inside the $n=7$ box is smaller than 150\,km\,s${^{-1}}$;
    %\item The number of undefined pixels in the neighboring points is under 50 \%.
    %\item The average over the neighboring points is also the highest (resp. lowest) along the circle.
    \item The sign of the velocity is the same as the sign of the median velocity over the neighboring points.
\end{itemize}
These criteria enabled us to remove most artifacts and to smooth sufficiently to get reliable results.  The minimal and maximal velocities $V_{min}$ and $V_{max}$ are thus computed for each ring. Figure~\ref{fig:minmax_all} displays the resulting map produced with $N=300$ concentric circles, $R_{max} = 1$~kpc and $R_{min}=0$. 

The same analysis was conducted on the second velocity component, defined with a flux ratio above 30 \% (as displayed in the bottom panels of Fig.~\ref{fig:kin}). Only the minimal points are computed for this second component, as the maxima are in patchy and noisy parts of the map.
We superpose these secondary velocities $V^{2nd}_{min}$ as green points in Fig.~\ref{fig:minmax_all}.

\subsubsection{Separate regions of the velocity map} \label{subsec:regions}
The extreme values of the velocity as a function of the galactocentric distance reveal well-defined independent structures in the velocity map (see Fig.~\ref{fig:minmax_all}). Firstly, this analysis stresses the very asymmetrical nature of the gas component of M31, detected both in the kinematics and in the structural characteristics. This asymmetry is mainly noticeable on both sides of the minor axis of the kinematics, as described in the following. 
\paragraph{Northeast region} 
This region hosts the maximal values of the velocity, which follow an almost-continuous line from the center to the edge of the map. The lines follow the winding visible at the center of the map. Around 0.4~kpc, the winding abruptly stops and the line roughly follows the axis of the position angle of Andromeda, around $PA=35^\circ$ from the northern axis (cf. Hyper-Linked Extragalactic Databases and Archives (HyperLEDA)). Such a structure is compatible with the signature of a rotating gaseous disk, whose position angle varies with the galactocentric distance. This variation induces a warped gaseous disk, similar to that depicted in \citet{Raj2021}. Moreover, the value of the velocity as a function of the galactocentric distance (right panel of Fig.~\ref{fig:minmax_all}) suggests a distorted rotation curve, hinting toward a variation of the inclination of the gaseous disk as well. We can already note at this stage that the few red points found at the eastern-most part of the map are not compatible with a warped gaseous disk, and may indicate non-circular motions. 

\begin{figure}[h]
    \centering
    \includegraphics[width=0.4\textwidth]{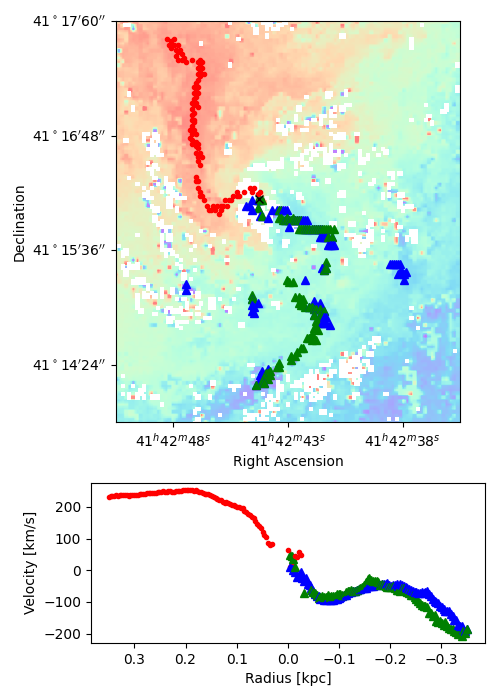}
    \caption{Zoom-in on the winding in the very center of the map. The analysis is conducted on $N=100$ radii from $R_{min}=0$ to $R_{max} = 0.35$~kpc. Top panel: $V_{min}$ (blue), $V^{2nd}_{min}$, and $V_{max}$ velocities superposed on the kinematic map, together with the maximal velocity (red bullets), minimal velocity (blue triangle) and minimal velocity of the second component (green triangle), for each radii.
    Bottom panel: Corresponding velocity values  as a function of the  galactocentric radius. The negative radii correspond to the distance of the points in the southwest region.}
    \label{fig:minmax_center}
\end{figure}

\paragraph{Southwest region}  % Southwestern region
On the contrary, this region hosts several distinct structures rather than an almost continuous line. One of those structures, near the center, is the counterpart of the winding seen in the northeast region as shown in Fig.~\ref{fig:minmax_center}. 
As seen in Fig.~\ref{fig:PA_infered}, the position angle of the red and blue symbols in the central 350~pc radius are roughly symmetrical. Moreover, the values of the velocity for the first and second velocity component (blue and green triangles) are in the same order of magnitude as their northeast counterpart, with a probable asymmetrical warp of the gaseous disk in this region.

Besides the winding pattern visible for galactocentric distances below 400~pc, the distribution of the minimal values on the map shows small segments, but no long and coherent structure emerges, neither in the primary nor in the secondary velocity component. This suggests disk tearing rather than disk warping, inducing discontinuity in the geometrical morphology of the gas in the southwest region. Beside the rift of the tearing, we expect double velocity components, as we would have the two gas structures that were teared apart really close to one another on the line of sight. This prediction is compatible with two bottom panels of Fig.~\ref{fig:kin}, which shows a small crescent of high velocity difference ($~-300$ km/s) where the rift would be. This could also account for the perturbed velocity dispersion pattern displayed in Fig.~\ref{fig:sig}.

\subsubsection{Constraints on the gas distribution}
The distribution of the extreme values extracted from the kinematic maps at different galactocentric distances suggests:
\begin{enumerate}
    \item a highly asymmetrical structure,
    \item a warped disk in the northeast region, with continuously varying position angle (and inclination),
    \item a teared disk in the southwest region, with discontinuously varying position angles (and inclination).
\end{enumerate}

\begin{figure}[h]
    \centering
    \includegraphics[width=0.3\textwidth]{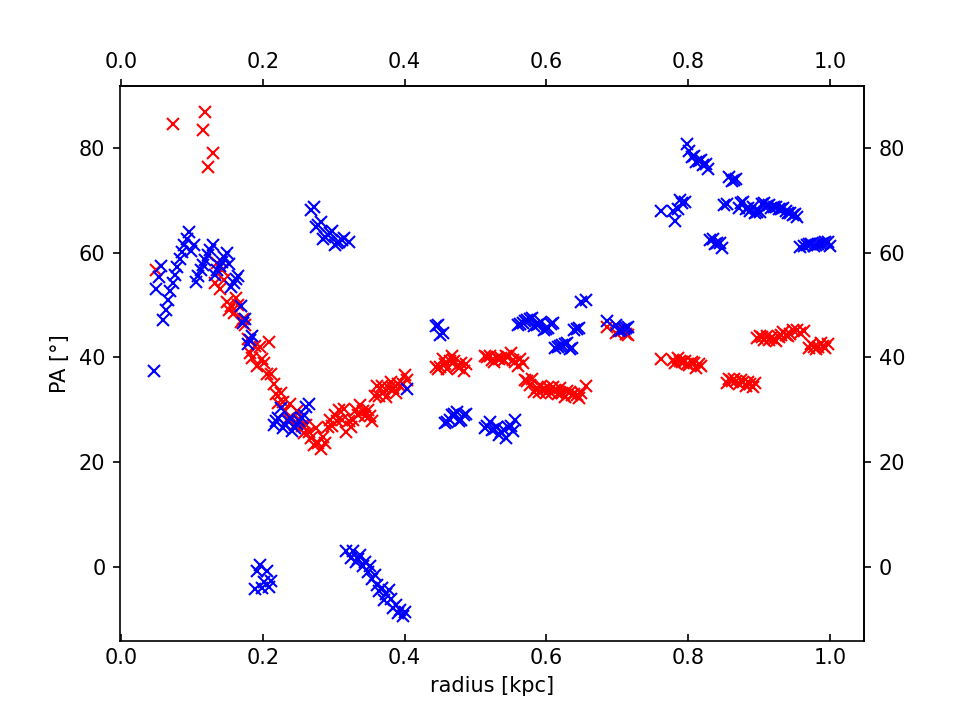}
    \caption{Position angle (PA) of the main kinematic features as a function of the galactocentric center, computed for the maximal values of the velocity in the northeast region (red) and the southwest region (blue). We distinguish the winding below 300~pc. Beyond 300~pc, the position angle slowly tends toward approximately 40$^\circ$.}
    \label{fig:PA_infered}
\end{figure}

In addition, we define the position angle (PA) as the angle between the northern axis and the direction of the maximal point from the center. Figure~\ref{fig:PA_infered} displays this PA value as a function of the galactocentric distance. While the redshifted points follow relatively regular pattern, the blueshifted points are irregular but for the inner 300~pc.
The value of the inclination angle is discussed in the subsequent section,
as it depends on the gravitational potential.

\section{Dynamical modeling}
\label{sect:dyn}
The goal of the present work is to model the ionized gas in the inner kpc of M31 and to reproduce the peculiar features displayed in its velocity field in the right panel of Fig.~\ref{fig:kin}.
Several disk components, with different inclinations and position angles will be considered, to take into account the warping and misalignment of the gas in the central perturbed kpc zone.
We identify four peculiar regions of interest:
\begin{enumerate}
    \item The winding of the velocity field at the very center of the map;
    \item The discontinuity of the velocity in the eastern part of the map, where we can see negative (blue) velocities superposed with positive (red) velocities (see zone between points A and B in Fig.~\ref{fig:minmax_all});
    \item A crescent-shaped region of high velocity (and high velocity gradient) in the southwest region;
    \item The main disk of M31, visible through projection in the northeast part of the map.
\end{enumerate}
As the gas content in the central kpc is weak in molecular \citep{Melchior2011,Melchior2013,Melchior2016,2017A&A...607L...7M,2019A&A...625A.148D}, in atomic \citep{Braun2009}, and in ionized gas \citep{1985ApJ...290..136J}, we assumed the gravitational potential to be dominated by stars. We study the kinematics of the structure \textit{via} static simulations of gas particles, located in a quasi-spherical potential. The positions of the particles are given by the geometry of the gas component, while their velocity vectors are derived from the total potential. We assume circular orbits corrected by an asymmetric drift and we choose to adapt our modeling inspired by \citet{Raj2021}. Their work presented a simulation of misaligned disks, characterized with different position angles and inclinations, spinning around the galaxy nucleus, and their torque-driven differential precession triggers warps in the inner galaxy. We thus consider the central part of M31 to be a set of misaligned circles, each one with its own position angle $\theta$ and inclination angle $i$. The angles $\theta$ and $i$ depend only on the distance $r$ from the center of the galaxy. \par
In this section, we describe the principle and the main features of the modeling. In Sect.~\ref{subsec:potentials}, we define the gravitational potential in the central region of M31. Then, in Sect. \ref{subsec:rho_v}, we derive the density and velocity fields. Finally, in Sect.~\ref{subsec:geom}, we describe the geometry of the gas in these fields.
We chose to use a symmetric modeling, and to adjust separately the NE and SW regions of the map.

\subsection{Potentials} \label{subsec:potentials}

The gravitational potential results from the superposition of the potentials produced by each massive structure composing the galaxy. 
We represent the spherical components (bulge, dark matter) by Plummer potentials, and the disks by Miyamoto-Nagai potentials \citep{Miya1975}. 
Contrary to previous models of the stellar component out to 6~kpc in radius \citep[e.g.,][]{Blana2017, Blana-Diaz2018, Saglia2018}, our goal is not to obtain a very precise model of the stellar bulge, but to model only the gaseous disk within 1~kpc, in terms of several components, warped and tilted. Our main point in the present paper is to focus on modeling the complex gas structure; for that we need an analytical model of the potential to derive easily frequencies and asymmetric drift, but not a perfect bulge model.

The potential $ \Phi_{Plummer}$ of a Plummer sphere of scale $R$, at a distance $r$ from its center, is given by:
\begin{equation}
    \Phi_{Plummer}(r) = -\frac{G M}{\sqrt{r^2 + R^2}}
,\end{equation}
where $M$ is the total mass of the component, and $G$ the gravitational constant.  For a disk of scale length, $R$, thickness, $h$, and total mass, $M$, the value of the Miyamoto-Nagai potential $\Phi_{MN}$, in cylindrical coordinates $(r,z)$, is:
\begin{equation}
    \Phi_{MN}(r,z) = -\frac{G M}{\sqrt{r^2 + (\sqrt{z^2+h^2} + R)^2}}
.\end{equation}
As a result, the spherical components have two parameters: their mass $M$ and scale length $R$, whereas the disks have three: their mass, $M$, scale length, $R$, and thickness, $h$. We take into account five structures of different sizes, shapes, and natures:

\begin{enumerate}
    \item The dark matter halo, characterized by $M_{DM}, \ R_{DM}$;
    \item The main stellar disk, characterized by $M_{SD}, \ R_{SD}$, and $h_{SD}$;
    \item The gaseous disk, characterized by $M_{GD}, \ R_{GD}$, and $h_{GD}$;
    \item A classical spherical galactic bulge, characterized by $M_{B}, \ R_{B}$;
    \item A nucleus, characterized by $M_{N}, \ R_{N}$.
\end{enumerate}

The total potential is the sum of these five potentials; hence, it is partly cylindrical and not completely spherical. However, the central kpc is dominated by the bulge, so the total potential can be considered almost spherical. This justifies our simplified assumption of circular orbits for the particles, depending only on the distance from the galactic center. 
Also, we did not take into account the inclination of the disk components when warped or tilted to compute their velocities, since the stellar disk brings a perturbation of less than 4\% on the rotational velocity (within the central kpc, as can be seen in Fig.~\ref{fig:rota}).
The nucleus component accounts for the mass of the three stellar disks discussed in \citet{Bender2005} as well as the black hole.

\subsection{Density and velocity fields} \label{subsec:rho_v}

\begin{figure}[h]
    \centering
    \includegraphics[width = 0.4\textwidth]{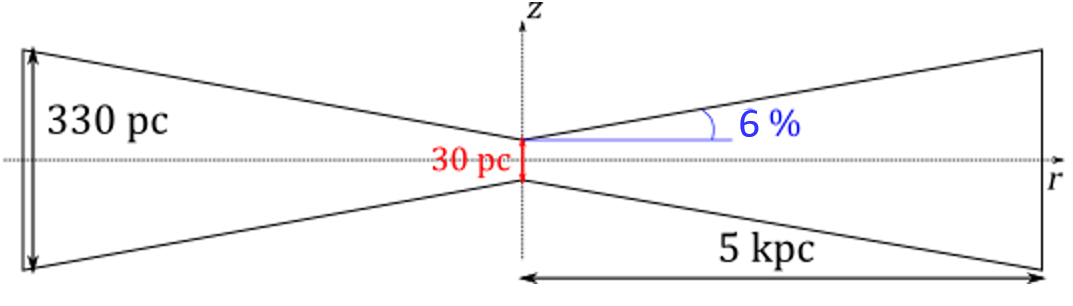}
    \caption{Cross-section of the gaseous disk along a diameter. The crossing of the $r$ and $z$ axis represents the center of the galaxy.}
    \label{fig:flare}
\end{figure}

To generate particles to model the central region of M31, we define the density and velocity fields. We assume for the gas the density $\rho(M)$ corresponding to a Miyamoto-Nagai disk, and we distribute $2.4\times 10^6$ gas particles inside a peculiar disk, whose thickness $H$ depends on the radius $r$, following: 
\begin{equation}
    H(r) = \alpha\times r + \beta 
    \label{eq:flare}
,\end{equation}
where $\alpha$ is fixed at 6\%, and $\beta$ is the thickness of the disk in the galactic center, fixed at $30$~pc. The radius of the disk in which the particles are generated is 5~kpc, and its maximal thickness is 330~pc (see Fig.~\ref{fig:flare}).

Each of the gas particle is given a velocity vector, whose norm is derived from the total potential $\Phi$ as followed:
\begin{equation}
    ||\overrightarrow{v}||(r) = \sqrt{r \frac{d\Phi}{dr}}
    \label{eq:velo}
.\end{equation}
We assumed circular orbits, with some epicyclic perturbations, accounting for the velocity dispersion, and therefore add an asymmetric drift. We adopt a Toomre parameter of $Q=1.3$, as discussed in \citet{Melchior2011}, and compute the radial, and azimuthal dispersions according to the epicyclic approximation, and the $z$ dispersion as a function of the plane thickness. To sum up, we have 2.4 million gas particles, each characterized by a coordinate vector $(x,y,z,v_x,v_y,v_z)$ in phase space.

The result of this distribution is a flared disk, whose velocity field would be a classical spider diagram. In this work, we assume that a warping and tilting of the disk can result in the peculiar velocity field we have observed. The mass of the gas component is known to be weak, which allows us to consider its morphological transformation as negligible with respect to the total potential.

\subsection{Geometry}
\label{subsec:geom}
We modeled the gas in the central part of the galaxy as a continuous set of concentric rings, whose radius range is from 0 to 5~kpc, to take into account the high inclination of the main disk. We generated 2.4 million particles, characterized by a set of six parameters $(x,y,z,v_x,v_y,v_z)$ inferred from the Eqs~\ref{eq:flare} and \ref{eq:velo}. 
To model a warped disk, each concentric circle is tilted and offset, to reproduce the kinematic geometry, according to adjustable parameters. Hence, a circle is characterized by a set of five parameters: $r$ its radius, $\theta$ its position angle, $i$ its inclination angle, $\delta N$ its offset in the northern direction, and $\delta E$ its offset in the eastern direction. Each of the 2.4 million particles and their velocity vector are rotated, then translated, to model the misalignment and the offset of the circles. After these transformations, the new coordinates are $(x',y',z',v'_x,v'_y,v'_z)$. The absolute value of the velocity vector (corresponding to the circular velocity) is not modified by the transformation. 
The galactocentric distance is modified only if there is an offset. The characteristic of the transformation depends only on the initial galactocentric distance: $r=\sqrt{x^2 + y^2 + z^2}$. 
We do not take into account the coordinate shifts to modify the corresponding velocities. This simplifying approximation is justified by the complexity of optimizing for too many parameters.

\begin{figure}[h]
    \centering
    \includegraphics[width=0.4\textwidth]{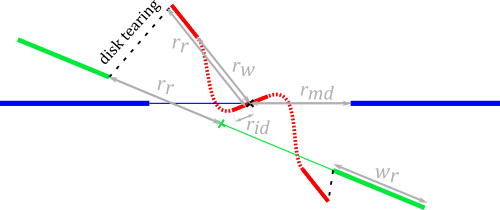}
    \caption{Schematic representation of the geometry of the gas in the central kpc of M31. The graph represents a simplified cross section of the gas disk in the sky plane containing the central black hole, represented by the black cross. The blue part is the main disk, aligned with the galactic disk of M31, the green part represents the ring and the red part shows the nuclear disk. The green cross represents the center of the ring, offset with respect to the nucleus. The dashed part of the red line corresponds to the warped disk, whereas the full lines are stiff components. The dashed black line points to where the disk is supposedly teared. We also show a number of parameters of the modeling: $r_w$ is the distance between the central black hole and the end of the warped region; $r_{md}$ indicates where the main disk begins; $r_r$ is the inner radius of the ring, as well as the end of the nuclear disk as they are supposed to be from a single structure that teared; $w_r$ is the width of the ring; and $r_{id}$ is the radius of the stiff disk inside the inner region. The substructures are not to scale for visibility reasons.}
    \label{fig:schem_geom}
\end{figure}

\begin{figure}[h]
    \centering
    \includegraphics[height = 8cm]{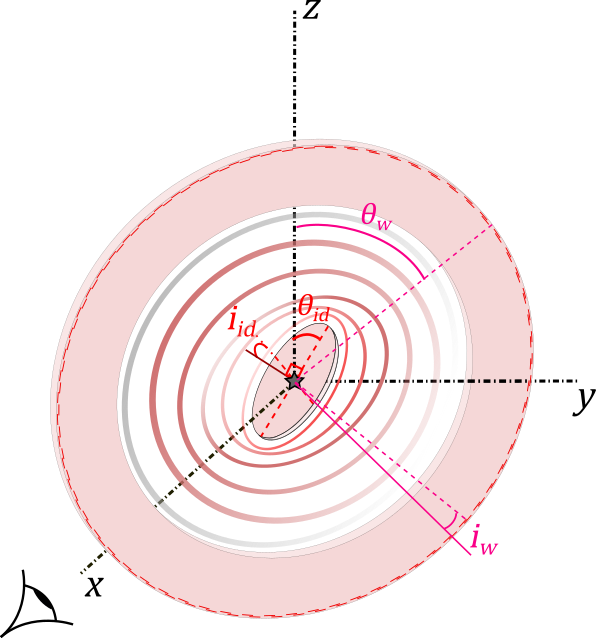}
    \caption{Schematic representation of the nuclear warped disk (corresponding to the red substructure in Fig.~\ref{fig:schem_geom}). The black star is the center of M31. The $x$ axis represents the line of sight, and the $yOz$ plane the plane of sky. The filled pink color refers to the inner nuclear disk, and also to the end of the warped region. The dashed lines represent two perpendicular diameters of the substructure. One of the two coincides to the major axis of the projection. The plain lines show where the second diameter would be if the substructure was not inclined (i.e., fully in the plane of sky). The successive circles in the middle are a simple representation of the warped region, defined as a set of misaligned circles with continuously varying inclination and position angles. For visibility reasons, the different parts of the substructure are not to scale with respect to each other, and only a few of the misaligned circles have been drawn. Bright red line at the outer bound of the substructure shows where the warped disk supposedly ripped from the 1~kpc ring.
    The nuclear warped disk is located inside the 1~kpc radius of M31, superposed with the $\sim$ 1~kpc inner ring and the main disk. }
    \label{fig:parameters1}
\end{figure}

We decompose the gaseous disk between three substructures, independent from one another and adjustable separately: (1) a main disk, belonging to the galactic plane, centered on the black hole; (2) an a warped nuclear disk, tilted with respect to the galactic plane, centered on the black hole; and (3) an inner ring, tilted with respect to the galactic plane, with an offset center. The nuclear disk presents a warped zone and is teared apart from the main disk. The disk tearing phenomenon happens with a geometry very similar to that described in \citet{Raj2021}, but for different physical reasons. The ring is also teared apart from the remainder of the gaseous disk. 
Apart from the nuclear disk, we consider the substructures to be stiff: inside the ring or inside the main disk, the values of the position angle, inclination angle, and offset are constants. A simplified schematics of the geometric modeling can be found in Figs.~\ref{fig:schem_geom} and~\ref{fig:parameters1}.

\subsubsection{Main disk} \label{subsubsec:md}

This disk has a hole in the middle, where we only see the inner and nuclear disk. The radius of the hole $r_{md}$ is an adjustable parameter of the modeling. The other parameters' values are fixed at the main orientation of the M31 disk: $\theta_{md} = 37^\circ$, $i_{md} = 77^\circ$, and $\delta E = \delta N = 0$. The thickness of the main disk follows the Eq.~\ref{eq:flare}. Even though the truncation radius we use (of 5~kpc) for the galactic disk is way beyond the field of view (approximately 1~kpc in radius), the ionized gas can still be seen in projection in the central area. 

\subsubsection{Ring} \label{subsubsec:r}

\begin{figure}[h]
    \centering
    \includegraphics[height = 8cm]{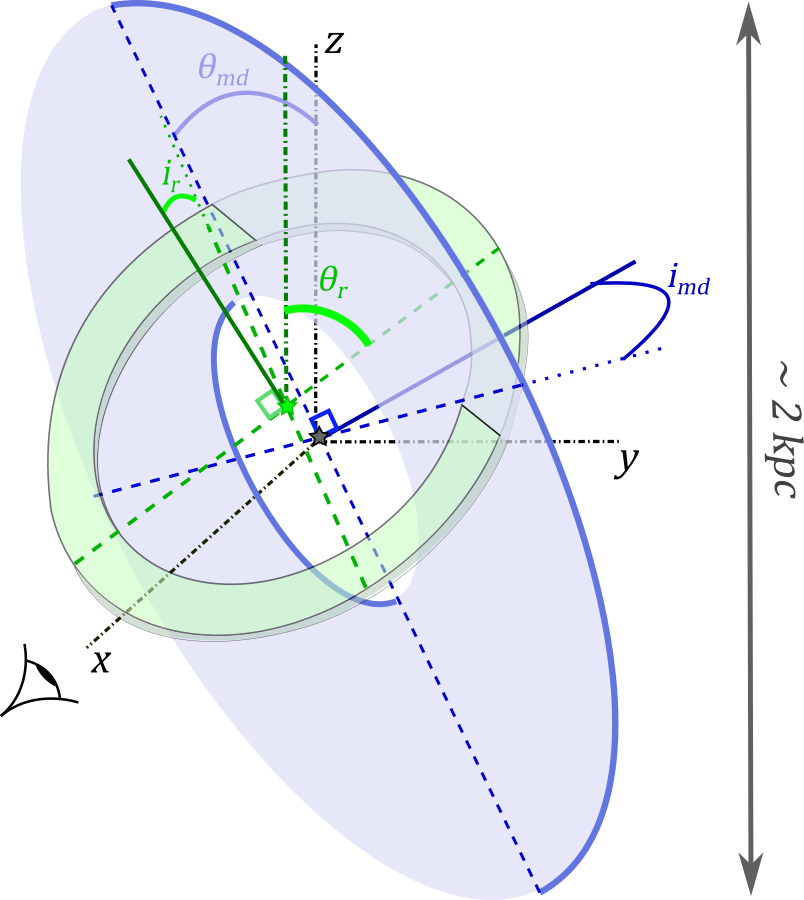}
    \caption{Schematic representation of the main disk and the inner ring (corresponding to the blue and green substructures in Fig.~\ref{fig:schem_geom}). The stars represent the center of the ring (in green), and the center of M31 (in black). In our modeling, we neglect the component of the offset parallel to the line of sight, and only consider the offset toward the east direction (here $-\vec{u_y}$) and north direction (here $\vec{u_z}$). The $x$ axis represents the line of sight, and the $yOz$ plane the plane of sky. The green dotted and dashed axis is parallel to the $z$ axis. In each color, the dashed lines represent two perpendicular diameters of the substructures. One of the two coincides to the major axis of the projection. The plain lines show where the second diameter would be if the substructure was not inclined (i.e., fully in the plane of sky). The angular parameters of the modeling are ($i_r$, $\theta_{r}$) and ($i_{md}$, $\theta_{md}$); the inclination and position angles of the ring and of the main disk.}
    \label{fig:main_ring}
\end{figure}

The inner ring is supposed to coincide with the dust ring seen in Spitzer maps and identified by \citet{Block2006}. We note that a schematic representation of this inner ring is displayed in Fig.~\ref{fig:main_ring}.
 It is tilted and offset with respect to the main disk, thus, it can be defined by six parameters: 
\begin{enumerate}
    \item $r_r$ the inner radius (see Fig.~\ref{fig:schem_geom});
    \item $w_r$ the width (see Fig.~\ref{fig:schem_geom});
    \item $\theta_r$ the position angle;
    \item $i_r$ the inclination;
    \item $\delta N_r$ the offset of the ring center in the northern direction;
    \item $\delta E_r$ the offset of the ring center in the eastern direction.
\end{enumerate}

\subsubsection{Warped nuclear disk} \label{subsubsec:id}

\begin{figure}[h]
\centering 
\includegraphics[width=0.4\textwidth]{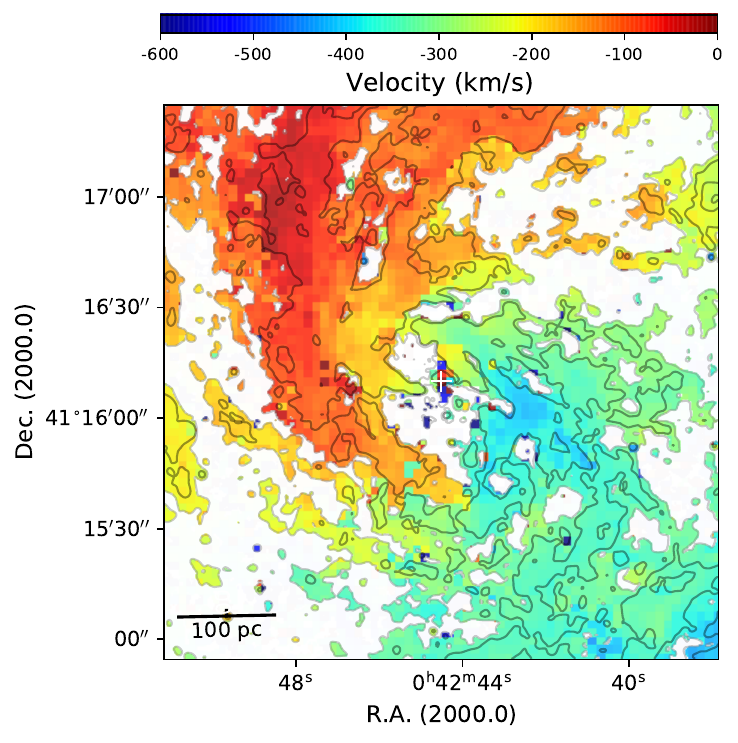}
\caption{Central 75$\times$75~arcsec$^2$ (285~pc $\times$285~pc) field of view of Andromeda. The white cross corresponds to the optical center. The intensity contours are the same as in the bottom panels of Fig.~\ref{fig:kin}.}
\label{fig:center}%
\end{figure}

The goal of the nuclear disk is to reproduce the winding at the very center of the velocity map, as presented in Fig.~\ref{fig:center}, as well as the discontinuity in the velocity profile as seen in the northeast part of the map (see Fig.~\ref{fig:kin}).
The varied geometrical angles are defined in Fig.~\ref{fig:parameters1}.

\section{Fitting the kinematics}
\label{sect:fitting}

The parameters for the modeling are adjusted to fit the velocity map shown in the top right panel of Fig.~\ref{fig:kin}. The modeling can be decomposed between the gravitational potential induced by the massive components of the galaxy, and the morphology of the gaseous disk immersed within this potential. Within 1~kpc scale, we expect the potential to be dominated by the stars, mainly apportioned in the bulge, the main stellar disk, and the nucleus.
The parameters of the potentials are needed to determine a theoretical rotation curve for the galaxy (Sect.~\ref{subsec:rota_curve}), which will then be used to infer the geometrical parameters of the gas (Sect.~\ref{subsec:geom_constraints}).

\subsection{Theoretical rotation curve} \label{subsec:rota_curve}

The theoretical rotation curve is entirely determined by the morphology of the massive structures (as explained in Sect.~\ref{subsec:potentials}). Several previous works give a good approximation of the majority of these parameters. 
The main stellar disk parameters $M_{SD}$, $R_{SD}$, and $h_{SD}$ are set according to \citet{2023AJ....166...80D}.
The dark matter halo parameters $M_{DM}$ and $R_{DM}$ are set according to \citet{Melchior2011}, as well as the parameters for the galactic bulge $M_{B}$ and $R_{B}$.
The nucleus parameters $M_{N}$ and $R_{N}$ are set according to \citet{Bender2005} and \citet{Bacon2001}.
Finally, we assume the mass of the gaseous disk to be 10\% of the stellar disk mass, and to extend to a 2~kpc-radius, with a maximal thickness of 200~pc.

In many of these works, the modeling chosen for the disks were exponential disks. Using a 2D curve fitting, we converted the scale and mass parameters to fit a Miyamoto-Nagai disk.
A summary of the chosen parameters is displayed in Table~\ref{tab:param}.

\begin{table}[h]
    \centering
    \caption{Physical parameters chosen to infer the theoretical rotation curve from the gravitational potentials.}
    \begin{tabular}{|c|c|c|c|c|}
    \hline
         &  Component&  $M$&  $R$& $h$ \\
 & & $[10^9 M_\odot]$& [kpc]& [kpc]\\
 \hline
         P&  Dark matter ($_{DM}$)&  270&  10&  --  \\
         MN&  Stellar disk ($_{SD}$)&  71&  2.5&  0.9\\
         MN&  Gaseous disk ($_{GD}$)&  7.1&  5.&  0.33\\
        P& Bulge ($_{B}$)& 16& 0.2& --\\
         P&  Nucleus ($_{N}$)&  0.14&  0.008&  -- \\ \hline
    \end{tabular} \vspace{0.2cm}
    \tablefoot{ The values are taken from literature \citep{Melchior2011, 2023AJ....166...80D, Bender2005, Bacon2001}. P stands for Plummer sphere and MN stands for Miyamoto-Nagai.}
    \label{tab:param}
\end{table}

\begin{figure}
    \centering
   \includegraphics[width=0.3\textwidth]{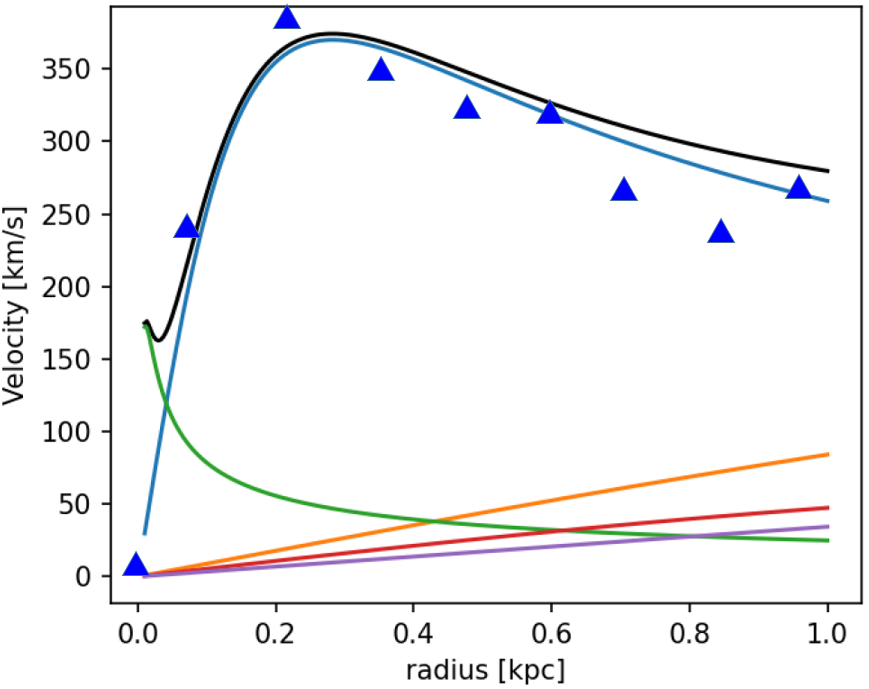}
    \caption{Theoretical rotation curve derived from Eq.~\ref{eq:velo}, using parameters from Table~\ref{tab:param}. The contributions are from the nucleus (green), the bulge (blue), the stellar disk (orange), the gaseous disk (red), and the dark matter halo (purple). The black line is the total rotation curve. The blue triangles correspond to the measured ionized gas velocities, averaged over the NE and SW side, and corrected from asymmetric drift.}
    \label{fig:rota}
\end{figure}

The theoretical rotation curve is computed using Eq.~\ref{eq:velo}. The contribution of each component is represented on Fig.~\ref{fig:rota}.
We computed the asymmetric drift, the difference between the rotational velocity $V_{rot}$ and the circular velocity $V_{circ}$, according to the following formula, adapted to the mid-plane of an axisymmetric rotating disk of density $\rho (R)$:
$$
V_{circ}-V_{rot} = \frac{\sigma_r^2}{2 V_{circ}} \, [\frac{\sigma_\theta^2}{\sigma_r^2}-1-
\frac{dln\rho\sigma_r^2}{dlnR}]
$$
where the radial and azimuthal velocity dispersions $\sigma_r$ and $\sigma_\theta$ are computed from the epicyclic approximation, and the Toomre criterion discussed in Sect.~\ref{subsec:rho_v}.

\subsection{Geometrical constraints} \label{subsec:geom_constraints}

The parameters of the main disk are fixed. We let the position angle and inclination of the inner ring to vary, and also the geometrical parameters of the nuclear warped disk.
Since the southwest region is affected by clear velocity perturbations, plausibly due to a large-scale shock, the geometrical parameters were essentially fitted on the northeast region.
A summary of the geometrical parameters can be found in Table~\ref{tab:geom}.

\begin{table}[h]
    \centering
    \caption{Geometrical parameters chosen for the warped nuclear 
    disk, inner ring, and main disk, in the fiducial model (scenario 3).}
    \begin{tabular}{|c|c|c|c|}
    \hline
  Parameter &   Warped &  Inner ring & Main disk\\
            &   nuclear disk  &      & \\
 \hline
         incl ($^\circ$) & 85&   48 & 35\\
         PA ($^\circ$)   & 65&  -27 & 77\\
         R (pc)         & 30&  900 & 400\\
         Width (pc)     &   &  300 & \\
         off-E (pc)     &  0&  -27 & 0\\
         off-N (pc)     &  0&  322 & 0\\ \hline
    \end{tabular} \vspace{0.2cm}
    \tablefoot{R is the inner radius for the ring and main disk, and outer radius for the warped nuclear disk.}
    \label{tab:geom}
\end{table}

\subsection{Fitting results} \label{subsec:fit-results}

\begin{figure}
\centering 
\includegraphics[width=0.2\textwidth]{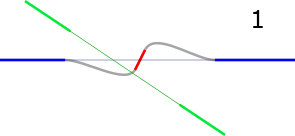}
\includegraphics[width=0.2\textwidth]{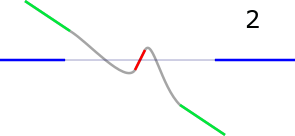}
\includegraphics[width=0.2\textwidth]{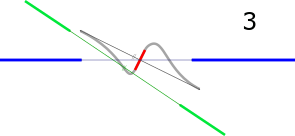}
\includegraphics[width=0.2\textwidth]{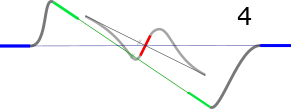}
\caption{Four scenarios proposed for the central region in M31, represented by their schematic transversal cut. Grey lines correspond to transition zones, blue lines to the main disk, red lines to the warped nuclear disk, and green lines to the inner ring. When two structures are decoupled, two distinct velocity components may be observed along the same line of sight, while when they are linked by a transition zone, there would be a single wide velocity component. 
Top-left: Scenario 1, where the nuclear and main disks are linked. The inner ring is decoupled, implying two velocity components along a line of sight. 
Top-right: Scenario 2, where the nuclear disk is linked to the inner ring. The main disk is decoupled, with possibly two velocity components on some line of sights. 
Bottom-left: Scenario 3, where the three structures are decoupled, implying distinct velocity components along a line of sight.
Bottom-right: Scenario 4, which is the same as scenario 3, but there is now a link between the main disk and inner ring, limiting the occurrence of distinct velocity components.}
\label{fig:4scenar}
\end{figure}

Four different scenarios have been explored for the complex geometry of the central kpc in M31. They are schematically described in Fig.~\ref{fig:4scenar}. They differ mainly by the degree of decoupling of the three different dynamical structures considered in projection in the central region. The corresponding geometrical parameters are listed in Tables~\ref{tab:geom} and \ref{tab:4scenar}.
 
\begin{figure}
    \centering
    \includegraphics[width=0.5\textwidth]{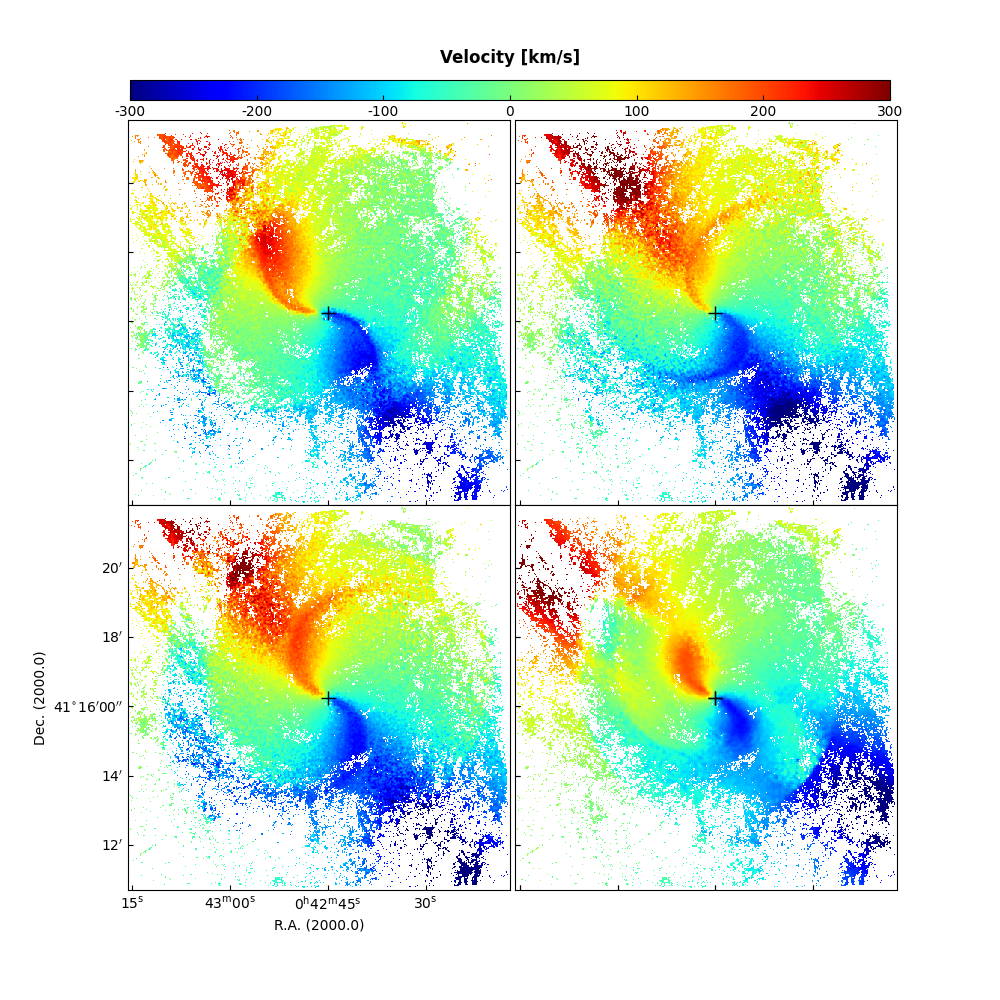}
    \caption{Velocity field obtained in the four scenarios described in Fig.~\ref{fig:4scenar}, in the same order. The maps have been masked with the velocity map of Fig.~\ref{fig:kin}; namely, we do not plot any velocity when there is no signal in the data.}
    \label{fig:4scenar-mom1}
\end{figure}

\begin{figure}
    \centering
    \includegraphics[width=0.5\textwidth]{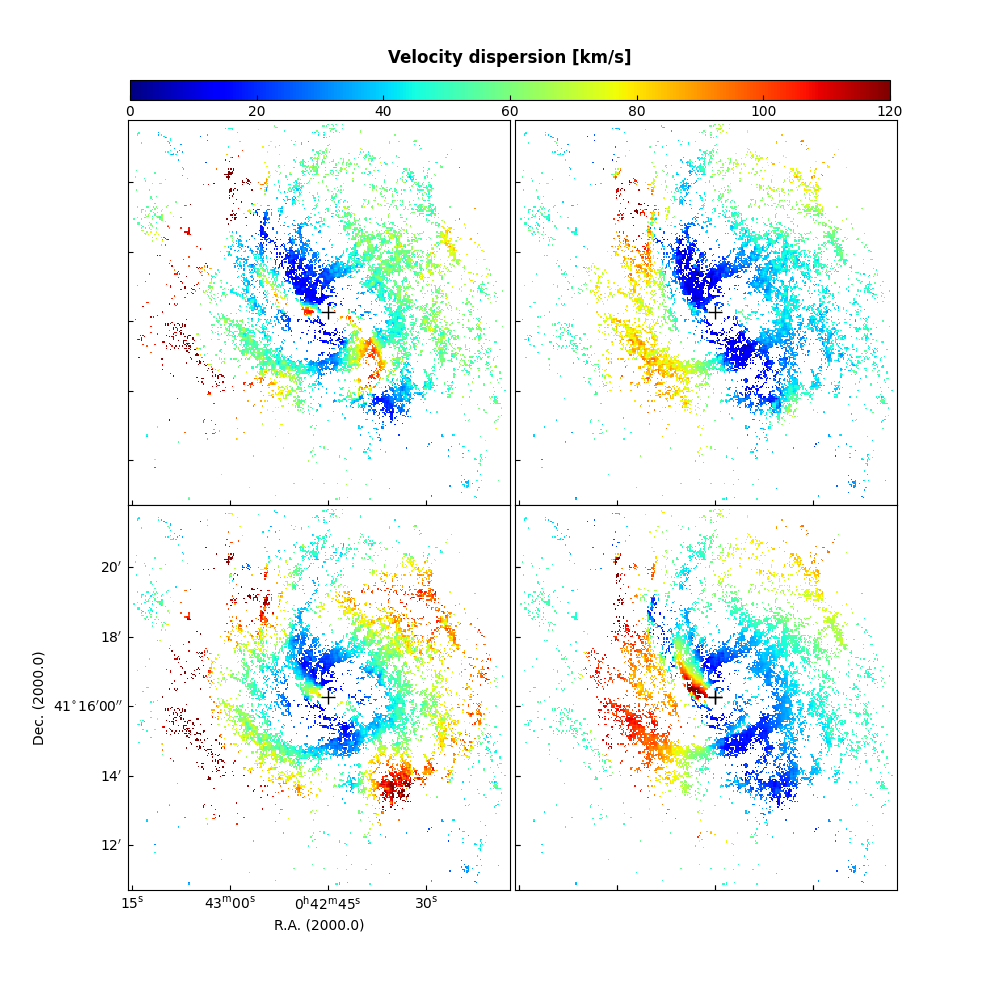}
    \caption{Same as Fig.~\ref{fig:4scenar-mom1} for the second moment of the cube. The maps have been masked with the velocity dispersion map of Fig.~\ref{fig:sig}; that is, we do not plot any velocity dispersion when there is no signal in the data}.
    \label{fig:4scenar-mom2}
\end{figure}

According to these parameters, we predict the velocity field in the four scenarios, represented on Fig.~\ref{fig:4scenar-mom1} and Fig.~\ref{fig:4scenar-mom2}, for the cube moments 1 and 2 respectively.
The fiducial model (scenario 3) is represented at the bottom left.
In scenario 1 (top left), the nuclear disk has an enhanced warp. The inclination of the ring is lower in scenario 2 (top right) and the radius of the inner warped disk smaller  in scenario 4 (bottom right).

\begin{table}[h]
    \centering
    \caption{Geometrical parameters chosen for the warped nuclear disk, inner ring, and main disk, in the additional scenarios.}
    \begin{tabular}{|c|c|c|c|}
    \hline
  Parameter &   Scenario 1 &  Scenario 2 & Scenario 4\\
 \hline
 Nuclear disk&    &   &  \\
      incl ($^\circ$) & 90&   43 & 85 \\
      PA ($^\circ$)   & 80&   70 & 65 \\
      R (pc)         & 70&   10 & 30 \\
  \hline
  Inner ring&    &   &  \\
      incl ($^\circ$) & 56&   40 &  45 \\
      PA ($^\circ$)   &-47&  -15 & -40 \\
      R (pc)         &860&  700 & 650 \\
      Width (pc)     &350&  300 & 400 \\
\hline
 Main disk&    &   &  \\
         R (pc)      &570&  700 & 1500\\
\hline
    \end{tabular} \vspace{0.2cm}
    \tablefoot{R is the inner radius for the ring and main disk, and outer radius for the warped nuclear disk. Only the varied parameters are displayed, the other fixed ones are the same as in Table~\ref{tab:geom}.}
    \label{tab:4scenar}
\end{table}

The velocity maps of Fig.~\ref{fig:4scenar-mom1} show that the velocity twist in the very center is reproduced in the four scenarios, thanks to the warped nuclear disk. 
Its amplitude is better fitted in scenarios 3 and 4.
The inner ring reproduces the blue-shifted region in the SE, and the best fit is scenario 3. The main disk is responsible for the red-shifted region in NE, and is optimally fit in scenarios 2 and 3. The high velocity dispersion in regions NW and SW is not perfectly reproduced in either scenario in Fig.~\ref{fig:4scenar-mom2}, but scenario 3 is approaching the closest. Globally, scenario 3 corresponds to the best qualitative model.

\section{Discussion}
\label{sect:discuss}

The recent kinematical data obtained with H$\alpha$ and [NII] lines on SITELLE in the central kpc of M31 have allowed us to build a fully sampled velocity field of the ionized gas. Combined with the more patchy molecular gas velocity field, obtained with the CO lines at IRAM-30m telescope, we identify three dynamical components, the main disk, a tilted ring, and a nuclear warped disk. The tilted ring is imposed by the discovery of two widely different velocity components (by $\sim$~260km/s) along the same line of sight, toward the center, around the minor axis \citep{Melchior2011, Melchior2016}. To account for the twisted velocity field around the very center (within 200~pc in radius; see Fig.~\ref{fig:kin}), we show that a warped nuclear disk, and then an off-centered ring propagating through a warped disk reproduce most of the observations. The signature of an almost face-on nuclear disk surrounded by a warped area is a good candidate to explain the winding found around the black hole of Andromeda. The extent of the winding, and its characteristic shape can be explained by the tearing of a warped zone between the tilted ring and the nuclear disk. The analysis of the velocity dispersion map shows that the highest dispersions are due to a superposition of distinct sub-structures, rather than a broadening of the velocity. By superposing the geometrical structures on the dispersion map, we see that a superposition of substructures induces dispersions between 80~\kms up to 120~\kms, a warped zone can induce a dispersion between 40 and 80~\kms where the warped region is edge on. The average dispersion for a simple disk structure is around 40~\kms. We can interpret in that way the observed velocity dispersion map, to deduce where we might find several velocities along the line of sight. This interpretation is corroborated by the double sinc fitting of the cube. 

We note that the warped structure of M31 has been reported by many authors in the past, namely in the HI gas \citep[e.g.,][]{Chemin2009} and references therein. Two warps have been seen, one in the outer parts, at radii larger than 18~kpc, and one in the nuclear disk, inside 6~kpc. \citet{Chemin2009} found that the nuclear region shows a "warp in the warp" and the gas is almost face-on here, according to rotation velocities, inclinations, and position angles. \citet{Ciardullo1988} also noted that the ionized gas distribution is more circular in the nuclear region, implying a face-on disk.
Over larger scales, an interpretation of the morphological and kinematical perturbations in the central 4~kpc of M31 has been proposed in terms of a bar, along with the accompanying non-circular motions and velocity jumps in shocks \citep{Blana-Diaz2018, Opitsch2018, Feng2022, Feng2024}.

Although the disk of M31 does not show evidence of a thin bar in the old stellar population, the triaxiality of the bulge has been recognized for a long time on red images \citep{Lindblad1956, Stark1977, Athanassoula2006, Beaton2007}. \citet{Stark1977} showed that the triaxiality of the bulge is witnessed by the twist of isophotes, which are misaligned with the disk PA, by an angle of $\sim$ 10$^\circ$. By modeling this morphology, \citet{Stark1977} estimated that a star in the disk at 5~kpc from the center would feel a tangential force typically about 5\% of the radial force. The triaxial bulge has axis ratios in the range 1:0.6:0.4, and extends to a radius of 2.6~kpc. The existence of an underlying bar in the disk is not obvious, due to the unfavorable inclination of the galaxy, but has been inferred from face-on models to be $\sim$ 4~kpc in radius \citep{Athanassoula2006}. However, the thin part of the bar, outside the boxy bulge, is not actually observed. The position angle of the near-infrared light outside the bulge has the same PA as the main disk \citep{Beaton2007}. The kinematics of the atomic gas shows some non-circular motions in the center, but nothing specific to the expected S-shape feature in the velocity field, due to a bar in the disk \citep{Braun1991, Chemin2009}. There is no bar signature in the HI morphology, but HI is depleted within 2~kpc. Instead there is a highly contrasted ring structure at radius 10~kpc, which appears superposed to some pieces of spiral arms. In the PV-diagram, it is possible to distinguish two HI rings, at 2.5 and 5~kpc, which are not centered, but lopsided \citep{Chemin2009}.

The ionized gas is not completely depleted in the center, and although the morphology does not show a bar response, but more a ring with apparent shocks, the kinematics reveal non-circular motions \citep{Opitsch2018}. In the center, the velocity field of the [OIII] line shows a kind of S-shape, mostly on the SE side. On the contrary, the stellar velocity field reveals regular rotation, showing no signature of bar and no S-shape feature.

\citet{Blana2017} provided a density model of the M31 bulge, composed of one-third nitial classical bulge (ICB) and two-thirds box-peanut bulge (BPB). The BPB alone is not sufficiently concentrated to fit the data. The triaxial bulge rotates with a pattern speed of $\Omega_p\sim$ 40~km/s/kpc, such that corotation is present at 5.8~kpc and OLR (outer Lindblad resonance) at 10.4~kpc. N-body simulations of only collisionless components (stars and dark matter) were performed and compared to the stellar data (morphology and velocities). The best fit is obtained from the morphology, position angle, boxiness, ellipticity, and asymmetry of the simulations within a radius if 4~kpc. Given their best fit model, and using the M2M (made-to-measure) technique to converge more accurately with the observations, \citet{Blana-Diaz2018} also fit the kinematical data obtained with the VIRUS-W IFU by \citet{Opitsch2018}. This allowed them to find the best mass-to-light ratio to account for the rotation curve and fit the stellar velocity field together with the velocity dispersion. The final model is a refinement and a better fit of the Gauss-Hermite kinematic parameters h3 and h4. More information could be provided via the comparison with the gas kinematics.

This approach was undertaken by \citet{Feng2022}, who analyzed the gas velocity field of the HI from \citet{Chemin2009} and the [OIII] from \citet{Opitsch2018}. They identified velocity jumps on a series of slits parallel to the minor axis. The width of the slits is 1.2’ = 274~pc, which is also the spacing between slit position. There are thus 17 slits from -9.6 to 9.6 arcmin from the minor axis. The position-velocity diagrams (PVD) are quite noisy in the [OIII] line, and when jumps are seen, they do not involve a large flux. Those correspond to the S-shape already remarked by \citet{Opitsch2018} in their velocity field. In the HI gas, the PVD are rather regular and expected for a smoothly rotating gas, which is not surprising, since the HI morphology does not show any signature of a bar. Hydrodynamical simulations are then carried out to illustrate the comparison, but not meant to match all the details of shocks. The result of the gas simulation has a morphology quite classical for a barred spiral galaxy, but very far from the M1 observations: there is a high central gas concentration in a nuclear disk with almost circular orbits, slightly perpendicular to the bar (x2 orbits), and then very narrow nuclear spirals, developing in an empty region, until a radius of $\sim$3~kpc. This morphology is not seen either in HI nor in the ionized gas. 

A more detailed analysis is provided by \citet{Feng2024}, with more adapted hydrodynamical simulations, using the same best fit model for the potential, obtained by the above M2M method. To obtain a better fit with the dynamical model, and to reproduce the position and amplitude of the identified shocks, they need a twice lower pattern speed of the bar, this time $\Omega_p\sim$ 20 km/s/kpc, such that corotation is now at 14~kpc. It is only if they allowed the nuclear disk to be tilted, with a more face-on orientation with respect to the large-scale disk, that the pattern speed could be increased to converge with the best fit of the stellar component. However, in this model, there is no reason for such a tilt. Again, the gas morphology in the model does not correspond to M31 observations, with a very high gas concentration in the center, with no lopsidedness. This disagreement of gas density is important, since the shocks are due in the model to gas with highly elliptical streamlines, spending a large time of their orbit in the center. However, this cannot be seen in the observations.

In summary, the bar model has serious problems when we are attempting to account for the gas morphology and kinematics in the central kpc of M31. Although there is a triaxial bulge, with significant pattern rotation, in addition to the classical bulge, which is able to reproduce the stellar kinematics, there might not be a thin bar present in the disk. One possibility is that M31 was a barred spiral galaxy before the encounter with an M32-like companion, as proposed by \citet{Block2006}. This almost head-on collision would have destroyed the thin bar in the disk, while propagating a ring wave, explaining the 10~kpc ring feature in the interstellar medium. The dynamically hotter triaxial bulge would have remained unperturbed after the encounter.

\section{Conclusion}
\label{sect:conclu}

We have built a fully sampled data cube of the ionized gas within a radius 1.2~kpc in the center of M31, through H$\alpha$ and [NII] mapped with SITELLE, at CFHT. Combined with the more patchy molecular gas velocity field, previously obtained with the CO lines at IRAM-30m telescope, and the dust photometry, we propose an interpretation of the complex central dynamics via a static modeling. We have introduced three dynamical components: the main disk with the inclination of 77$^\circ$ and PA= 37$^\circ$, a tilted ring, required to explain spectra with double-velocity components, and a nuclear warped disk. The mass model of the central kpc is dominated mainly by the bulge and the nuclear disk, although we added the main disk and the dark matter halo to fit the rotation curve. The kinematics of the ionized and molecular gas is then computed in this potential and the velocity field confronted with the observations. The tilted ring component is allowed to be off-centered, with a spatial excursion reproducing the observations. The morphology and orientation angles of the tilted ring and warped nuclear disk were optimized to obtain the best qualitative compatibility with the observations. Several close configurations are possible and the complex dynamics of the central region is thought to be the result of a recent head-on collision with a M-32 like galaxy, as previously proposed by \citet{Block2006}. The dynamical perturbations observed correspond to the nuclear disk tilted and warped in the collision which is now settling down back to equilibrium, keeping some lopsidedness, while the m=1 waves are damping down. The central kpc of M31 is still embedded in a larger triaxial box-peanut-shaped bulge, as a remnant of the pre-collision bar. However, the gas disk within 1-2~kpc is too perturbed to show any bar signature.

\begin{acknowledgements}
We are most grateful to the anonymous referee for the very constructive comments that helped us to substantially improve the manuscript. This work is based on observations carried out under project numbers 067-11 and 221-11 with the IRAM-30m telescope. IRAM is supported by INSU/CNRS (France), MPG (Germany), and IGN (Spain). This paper is based on observations obtained with SITELLE, a joint project of Universit{\'e} Laval, ABB, Universit{\'e} de Montr{\'e}al, and the Canada-France-Hawaii Telescope (CFHT) which is operated by the National Research Council (NRC) of Canada, the Institut National des Science de l'Univers of the Centre National de la Recherche Scientifique (CNRS) of France, and the University of Hawaii. The authors wish to recognize and acknowledge the very significant cultural role that the summit of Mauna Kea has always had within the indigenous Hawaiian community. 
\end{acknowledgements}

\bibliographystyle{aa}
\bibliography{final}

\end{document}